\documentclass[seceq]{ptptex}

\usepackage{graphicx}
\usepackage{wrapft}

\preprintnumber[3cm]{%<-- [..]: optional width of preprint # column.
NAOJ-Th-Ap 2003,No.19\\PTPTeX ver.0.8\\ February, 2003}
\newcommand{\MR}{{\rm I\!R}}
\title{%        %You can use \\ for explicit line-break
Gauge Invariant Variables\\ in Two-Parameter Nonlinear Perturbations
}

\author{%       %Use \scshape  for the family name
  Kouji {\sc Nakamura}\footnote{E-mail: kouchan@th.nao.ac.jp} 
}

\inst{%         %Affiliation, neglected when [addenda] or [errata]
  Division of Theoretical Astrophysics,
  National Astronomical Observatory, \\
  Tokyo, Mitaka 181-8588, Japan
}

\recdate{March 24, 2003}%            %Editorial Office will fill in this.

\abst{%         %this abstract is neglected when [addenda] or [errata]
The procedure to find gauge invariant variables for two-parameter
nonlinear perturbations in general relativity is considered.
For each order metric perturbation, we define the variable
which is defined by the appropriate combination with lower order
metric perturbations.
Under the gauge transformation, this variable is transformed in the
manner similar to the gauge transformation of the linear order
metric perturbation.
We confirm this up to third order.
This implies that gauge invariant variables for higher order
metric perturbations can be found by using a procedure similar
to that for linear order metric perturbations.
We also derive gauge invariant combinations for the perturbation of
an arbitrary physical variable, other than the spacetime metric,
up to third order.
}

\begin{document}

\maketitle

%%%%%%%%%%%%%%%%%%%%%%%%%%%%%%%%%%%%%%%%%%%%%%%%%%%%%%%%%%%%%%%%%%%%%%
\section{Introduction}
%%%%%%%%%%%%%%%%%%%%%%%%%%%%%%%%%%%%%%%%%%%%%%%%%%%%%%%%%%%%%%%%%%%%%%

The perturbative approach is one of the popular techniques
to investigate physical systems. 
In particular, this approach is powerful when the
construction of exactly soluble models is difficult.
In general relativity, exact solutions to the Einstein
equation are most often too idealized to properly represent
the realm of natural phenomena, though many exact solutions
are known\cite{Exact-solution}.
Here the constructing perturbative solutions around appropriate
exact solutions is a useful approach to investigate realistic
situations.
Cosmological perturbation
theory\cite{Bardeen1980}\tocite{Mukhanov-Feldman-Brandenberger}
is now the most commonly used technique, and perturbations of
black holes and stars have been widely studied to obtain
descriptions of the gravitational radiation emitted from
them.\cite{Chadrasekhar-book}\tocite{Kokkotas-quasi-normal} 
These recent perturbative analyses have been extended to
second-order perturbations, but in many cases, these treatments
employ an expansion in a single parameter.

%************************************************

In some physical applications, it is convenient to introduce two (or
more) infinitesimal perturbation parameters to elucidate the
physical meaning of the perturbations.
One typical example is the study of perturbations of rotating
stars.\cite{Kojima}\tocite{Stergioulas}
No exact analytic stationary axisymmetric solution describing
rotating stars has yet been obtained, at least for reasonably
interesting equations of state.
To treat rotating stars, perturbative analyses employing the
``slow rotating approximation'' are commonly used.
In this approach, the background is a non-rotating star,
i.e. spherically symmetric star, and two small parameters, $\lambda$ 
and $\epsilon$, corresponding to the pulsation amplitude and the
rotation parameter, are introduced.
The pulsation amplitude is given by the amplitude of the metric
perturbation, and the rotation parameter is given by
$\epsilon=\Omega/\sqrt{GM/R^{3}}$, where $\Omega$ is the uniform
angular velocity, and $M$ and $R$ are the mass and the radius of
the non-rotating star, respectively.
In this approach, the first order in $\epsilon$ describes frame
dragging effects, with the star actually remaining spherical,
and $\epsilon^{2}$ terms describe the effects of rotation on the 
fluid\cite{Hartle}. 
Because the mass-shedding limit corresponds to $\epsilon\sim 1$,
this approximation is valid for angular velocity $\Omega$ much smaller
than the mass-shedding limit, and this approximation has also
been used recently in the study of the instability in rotating stars.
(See the review paper by Stergioulas\cite{Stergioulas} and
references therein). 
This example shows that the two-parameter perturbation theory is
interesting from the viewpoint not only of mathematical physics but 
also of its applications. 
There are many astrophysical situations that should be analyzed
using multi-parameter perturbation theory.

%************************************************

In spite of these efforts, classical studies in the literature
have not been analyzed taking care of the full gauge dependence
and gauge invariance of the non-linear perturbation theory.
For example, the delicate treatment of gauge freedom is necessary
when we evaluate boundary conditions at the surface of the matter
distribution and the perturbative displacement of the surface 
(for example, see Refs.~24)).
An implicit fundamental assumption in relativistic perturbation theory
is that there exists a parametric family of spacetimes such that the
perturbative formalism is built as a Taylor expansion in this family
around a background.
The perturbations are then defined as the derivative terms of this
series, evaluated on this background\cite{Wald-book}.
To carry out this evaluation, we must identify the points on the
background spacetime and those on a physical spacetime that we
attempt to describe as a perturbation of the background spacetime. 
This choice of the identification map is usually called the gauge
choice\cite{J.M.Stewart-M.Walker11974}. 
The important point is that this identification is not unique, i.e.
there is a degree of freedom involved in the choice of this
identification map. 
This is the gauge freedom in the perturbation theory.
Clearly, this consists of more than the usual assignment of
coordinate labels to points of a single spacetime. 
Further, the Einstein equation does not determine this gauge
freedom, and we must fix this gauge freedom by hand or extract
the gauge invariant part of the perturbations (for example, see
Ref.~13)).
This problem does not arise when this gauge freedom is
completely fixed and if a change of the gauge is not necessary
to analyze or interpret the physical meanings of the results.
Otherwise, this problem always arises.
Therefore, it is important to clarify the gauge transformation
rules of physical variables and the concept of gauge invariance.

%************************************************

In this paper, we present the procedure to define gauge invariant
variables in the two-parameter nonlinear perturbation theory.
Recently, Bruni and coworkers derived the gauge transformations
and introduced the concept of gauge invariance in the
two-parameter nonlinear spacetime perturbation
theory\cite{Bruni-Gualtieri-Sopuerta}.
They derived explicit gauge transformation rules up to fourth
order, i.e., including any term $\lambda^{k}\epsilon^{k'}$ with 
$k+k'\leq 4$.
We follow their ideas in this paper.
Although we keep in mind the above mentioned practical examples,
we do not make any specific assumption regarding the background
spacetime and the physical meaning of the two-parameter family.
Instead, we assume the existence of a procedure to determine the
gauge invariant variables at linear order.
For each order metric perturbation, we define the variable
which is defined by the appropriate combination with lower order
metric perturbations.
Under the gauge transformation, this variable is transformed in
the manner similar to the gauge transformation of linear order
metric perturbation.
We confirm this up to third order.
This implies that we can always find gauge invariant variables
for higher order metric and matter perturbations because we have
a procedure to determine gauge invariant variables of linear
perturbations as we assumed.

%************************************************

Because we make no assumption concerning the background
spacetime, our procedure is applicable to various situations.
Further, we note that gauge freedom always exists in the
perturbation of theories in which we impose general covariance.
Therefore, our procedure is applicable not only to general
relativity but also to any theory in which general covariance
is imposed.
However, we cannot treat the situation in which the change of
the differential structure arises due to the perturbations.
We also note that the procedure developed here has already been
applied to clarify the oscillatory behavior of a gravitating
Nambu-Goto string\cite{kouchan-string-flat} in which it is
crucial to distinguish the gauge freedom of the perturbations
and the motion of the string.
Through such considerations, we have already confirmed that the
procedure we study here is applicable in a specific case.

%************************************************

The organization of this paper is as follows.
In \S\ref{sec:Taylor}, we present the necessary mathematical
tools, deriving Taylor expansion formulae for two-parameter
families of diffeomorphisms.
In \S\ref{sec:gauge-trans}, we set up an appropriate geometrical 
description of the gauge dependence in two-parameter families of
spacetimes and derive gauge transformation rules for the
perturbations. 
In \S\ref{sec:gauge-invariant}, the procedure to determine the
gauge invariant variables of nonlinear perturbations is described.
The final section, \S\ref{sec:Summary-Discussions}, is devoted
to summary and discussions. 
Sections \ref{sec:Taylor} and \ref{sec:gauge-trans} consists
largely of a review of the work of Bruni et
al.\cite{Bruni-Gualtieri-Sopuerta}, 
which is referred to as BGS2003 in the present paper.
However, these sections include some additional explanations that
are not given in BGS2003.
In particular, we note that the representation of the Taylor
expansion given in this paper is simpler but equivalent to that
given in BGS2003.
We employ the notation of BGS2003 and also use the abstract
index notation\cite{Wald-book}.

%************************************************

%%%%%%%%%%%%%%%%%%%%%%%%%%%%%%%%%%%%%%%%%%%%%%%%%%%%%%%%%%%%%%%%%%%%%
\section{Taylor expansion of the two parameter diffeomorphisms}
\label{sec:Taylor}
%%%%%%%%%%%%%%%%%%%%%%%%%%%%%%%%%%%%%%%%%%%%%%%%%%%%%%%%%%%%%%%%%%%%%

%************************************************

Perturbation theories on a manifold are usually based on a Taylor
expansion on an extended manifold of the original manifold.
Taylor expansions provide an approximation of the value of a
quantity at some point in terms of its value and the values of
its derivative, at another point. 
Here, a Taylor expansion of tensorial quantities can only be defined
in terms of a mapping between tensors at different points of the
manifold under consideration. 
This implies that a two-parameter perturbation theory on a manifold
requires a Taylor expansion of such a mapping given by a
two-parameter family of diffeomorphisms on the manifold.
In this section, we review the Taylor expansion of two-parameter 
diffeomorphisms developed in BGS2003, with some modifications
to clarify the essence of their idea.

%************************************************

Given a differentiable manifold ${\cal M}$, we consider a family
of diffeomorphisms $\Phi_{\lambda,\epsilon}$ characterized by
two parameters on ${\cal M}$, $\lambda$ and $\epsilon$: 
\begin{eqnarray}
  \Phi_{\lambda,\epsilon} : 
  {\cal M}\times\MR^{2} &\rightarrow& {\cal M}\times\MR^{2}
  \nonumber\\
  (p,\lambda,\epsilon) &\mapsto& (\Phi_{\lambda,\epsilon}(p),\lambda,\epsilon).
\end{eqnarray}
As emphasized by Bruni et al.\cite{M.Bruni-S.Soonego-CQG1997},
the diffeomorphisms $\Phi_{\lambda,\epsilon}$ do not form a group in
the form 
$\Phi_{\lambda_{1},\epsilon_{1}}\circ\Phi_{\lambda_{2},\epsilon_{2}} 
= \Phi_{\lambda_{1}+\lambda_{2},\epsilon_{1}+\epsilon_{2}}$
for all $\lambda_{i},\epsilon_{j}\in\MR$ ($i=1,2$).
This differs from the usual situation for exponential
maps\cite{BruhatMoretteBleick}.
In the generic case, we must keep in mind the fact that 
\begin{equation}
  \label{eq:Phi-not-a-group-by-parameter}
  \Phi_{\lambda_{1},\epsilon_{1}}\circ\Phi_{\lambda_{2},\epsilon_{2}} 
  \neq \Phi_{\lambda_{1}+\lambda_{2},\epsilon_{1}+\epsilon_{2}}.
\end{equation}
This means that $\Phi_{\lambda,\epsilon}$, in general, cannot be
decomposed into the form
$\Phi_{\lambda,\epsilon}=\Phi_{0,\epsilon}\circ\Phi_{\lambda,0}$,
where both $\Phi_{0,\epsilon}$ and $\Phi_{\lambda,0}$ are
one-parameter families of diffeomorphisms.
Hence, in the derivation of the representation of the Taylor expansion
of the pull-back $\Phi_{\lambda,\epsilon}^{*}$ of
$\Phi_{\lambda,\epsilon}$, we cannot use the representation of the
Taylor expansion of the pull-back of the one-parameter family of
diffeomorphisms\cite{M.Bruni-S.Soonego-CQG1997,S.Soonego-M.Bruni1998}.
The Taylor expansion based on the usual exponential maps is realized as
a special case of the representation derived here.

%************************************************

The simple algebraic properties of the coefficients of the
Taylor expansion of $\Phi^{*}_{\lambda,\epsilon} Q$ for an
arbitrary tensor field $Q$ leads to their representation in
terms of suitable Lie derivatives.
We start from the formal expression of the Taylor expansion of
the pull-back $\Phi^{*}_{\lambda,\epsilon} Q$, which is
given by
\begin{equation}
  \Phi^{*}_{\lambda,\epsilon} Q = \sum^{\infty}_{k,k'=0}
  \frac{\lambda^{k}\epsilon^{k'}}{k!k'!} 
  \left\{
    \frac{\partial^{k+k'}}{\partial\lambda^{k}\partial\epsilon^{k'}} 
    \Phi^{*}_{\lambda,\epsilon} Q
  \right\}_{\lambda=\epsilon=0}.
  \label{eq:formal-Taylor-expansion}
\end{equation}
As the properties of the coefficients of the Taylor expansion
(\ref{eq:formal-Taylor-expansion}), we stipulate that the operators
$\partial/\partial\lambda$ and $\partial/\partial\epsilon$ in the
bracket $\left\{*\right\}_{\lambda=\epsilon=0}$ in
Eq.~(\ref{eq:formal-Taylor-expansion}) are not symbolic notation
but, rather, the usual partial differential operators on $\MR^{2}$.
The representation of this Taylor expansion in terms of the Lie
derivatives is explicitly derived in Appendix
\ref{sec:taylor-derivation}. 
We note that the Leibniz rule plays a key role in the derivation
of the representation of the Taylor expansion
(\ref{eq:formal-Taylor-expansion}).

%************************************************\\

In this paper, we present the expansion of the pull-back 
$\Phi^{*}_{\lambda,\epsilon}Q$ to order $\lambda^{k}\epsilon^{k'}$
with $k+k'=3$ in terms of suitable Lie derivatives. 
For this purpose, we introduce the following set of operators 
${\cal L}_{(p,q)}$, where $p$ and $q$ are integers, on an
arbitrary tensor field $Q$: 
\begin{eqnarray}
  \label{eq:Bruni-14}
  {\cal L}_{(1,0)}Q
  &:=&
  \left\{ 
    \frac{\partial}{\partial\lambda}\Phi^{*}_{\lambda,\epsilon}Q
  \right\}_{\lambda=\epsilon=0} 
  , \\  
  \label{eq:Bruni-15}
  {\cal L}_{(0,1)}Q
  &:=&
  \left\{ 
    \frac{\partial}{\partial\epsilon}\Phi^{*}_{\lambda,\epsilon}Q
  \right\}_{\lambda=\epsilon=0}
  , \\  
  \label{eq:Bruni-16}
  {\cal L}_{(2,0)}Q
  &:=&
  \left\{ 
    \frac{\partial^{2}}{\partial\lambda^{2}}\Phi^{*}_{\lambda,\epsilon}Q
  \right\}_{\lambda=\epsilon=0} - {\cal L}^{2}_{(1,0)}Q
  , \\  
  \label{eq:Bruni-17}
  {\cal L}_{(1,1)} Q
  &:=& 
  \left\{ 
    \frac{\partial^{2}}{\partial\lambda\partial\epsilon} 
    \Phi^{*}_{\lambda,\epsilon}Q
  \right\}_{\lambda=\epsilon=0} 
  - \frac{1}{2} \left(
    {\cal L}_{(1,0)}{\cal L}_{(0,1)} 
    + {\cal L}_{(0,1)}{\cal L}_{(1,0)}
  \right)Q
  , \\ 
  \label{eq:Bruni-18}
  {\cal L}_{(0,2)}Q
  &:=& 
  \left\{ 
    \frac{\partial^{2}}{\partial\epsilon^{2}}\Phi^{*}_{\lambda,\epsilon}Q
  \right\}_{\lambda=\epsilon=0} 
  - {\cal L}^{2}_{(0,1)}Q
  , \\
  \label{eq:Bruni-19}
  {\cal L}_{(3,0)}Q 
  &:=& 
  \left\{ 
    \frac{\partial^{3}}{\partial\lambda^{3}}\Phi^{*}_{\lambda,\epsilon}Q
  \right\}_{\lambda=\epsilon=0} 
  - 3 {\cal L}_{(1,0)} {\cal L}_{(2,0)}Q 
  - {\cal L}^{3}_{(1,0)}Q
  , \\  
  {\cal L}_{(2,1)}Q
  &:=&
  \left\{ 
    \frac{\partial^{3}}{\partial\lambda^{2}\partial\epsilon}
    \Phi^{*}_{\lambda,\epsilon}Q
  \right\}_{\lambda=\epsilon=0} 
  - 2 {\cal L}_{(1,0)} {\cal L}_{(1,1)}Q 
  - {\cal L}_{(0,1)} {\cal L}_{(2,0)}Q 
  \nonumber\\
  && \quad
  - {\cal L}_{(1,0)} {\cal L}_{(0,1)} {\cal L}_{(1,0)} Q 
  \label{eq:Bruni-20}
  , \\
  {\cal L}_{(1,2)}Q
  &:=&
  \left\{ 
    \frac{\partial^{3}}{\partial\lambda\partial\epsilon^{2}}
    \Phi^{*}_{\lambda,\epsilon}Q
  \right\}_{\lambda=\epsilon=0} 
  - 2 {\cal L}_{(0,1)} {\cal L}_{(1,1)}Q 
  - {\cal L}_{(1,0)} {\cal L}_{(0,2)}Q 
  \nonumber\\
  && \quad
  - {\cal L}_{(0,1)} {\cal L}_{(1,0)} {\cal L}_{(0,1)} Q 
  \label{eq:Bruni-21}
  , \\
  {\cal L}_{(0,3)}Q
  &:=&
  \left\{ 
    \frac{\partial^{3}}{\partial\epsilon^{3}}\Phi^{*}_{\lambda,\epsilon}Q
  \right\}_{\lambda=\epsilon=0} 
  - 3 {\cal L}_{(0,1)} {\cal L}_{(0,2)}Q 
  - {\cal L}^{3}_{(0,1)}Q.
  \label{eq:Bruni-22}
\end{eqnarray}
As shown in Appendix \ref{sec:taylor-derivation}, the
above operators ${\cal L}_{(p,q)}$ are linear and satisfy the 
Leibnitz rule, and hence they are derivative operators.
Because the pull-back $\Phi^{*}_{\lambda,\epsilon}$ of a
diffeomorphism $\Phi_{\lambda,\epsilon}$ commutes with
contractions and the exterior
derivative\cite{Thirring-book,Kobayashi-Nomizu}, the operators
${\cal L}_{(p,q)}$ also commute with any contraction and
exterior derivative. 
Therefore, for each of them, there is a vector field
$\xi_{(p,q)}^{a}$ such that 
\begin{equation}
  {\pounds}_{\xi_{(p,q)}}Q := {\cal L}_{(p,q)}Q
  \label{eq:Lie-calL-rela}
\end{equation}
for $p,q=0,1,2,3$.

%************************************************\\

Using the Lie derivative (\ref{eq:Lie-calL-rela}), we can express the
Taylor expansion (\ref{eq:formal-Taylor-expansion}) of the pull-back
$\Phi_{\lambda,\epsilon}^{*}$ of $\Phi_{\lambda,\epsilon}$ in
terms of the Lie derivatives associated with the vector fields
$\xi_{(p,q)}^{a}$ (\ref{eq:Lie-calL-rela}) up to third order in
$\lambda$ and $\epsilon$:
\begin{eqnarray}
  \Phi^{*}_{\lambda,\epsilon}Q &=& Q
  + \lambda {\pounds}_{\xi_{(1,0)}}Q
  + \epsilon {\pounds}_{\xi_{(0,1)}}Q
  \nonumber\\
  && \quad
  + \frac{\lambda^{2}}{2} \left\{{\pounds}_{\xi_{(2,0)}} + 
    {\pounds}_{\xi_{(1,0)}}^{2}\right\} Q
  \nonumber\\
  && \quad\quad
  + \lambda\epsilon \left\{ {\pounds}_{\xi_{(1,1)}} 
    + \frac{1}{2} {\pounds}_{\xi_{(1,0)}} {\pounds}_{\xi_{(0,1)}} 
    + \frac{1}{2} {\pounds}_{\xi_{(0,1)}} {\pounds}_{\xi_{(1,0)}} 
  \right\} Q
  \nonumber\\
  && \quad\quad
  + \frac{\epsilon^{2}}{2} \left\{{\pounds}_{\xi_{(0,2)}} + 
    {\pounds}_{\xi_{(0,1)}}^{2}\right\} Q
  \nonumber\\
  && \quad
  + \frac{\lambda^{3}}{6} \left\{
    {\pounds}_{\xi_{(3,0)}} + 3 {\pounds}_{\xi_{(1,0)}} {\pounds}_{\xi_{(2,0)}}
    + {\pounds}_{\xi_{(1,0)}}^{3}
  \right\} Q
  \nonumber\\
  && \quad\quad
  + \frac{\lambda^{2}\epsilon}{2} \left\{
    {\pounds}_{\xi_{(2,1)}}
    + 2 {\pounds}_{\xi_{(1,0)}} {\pounds}_{\xi_{(1,1)}} 
    + {\pounds}_{\xi_{(0,1)}} {\pounds}_{\xi_{(2,0)}}
    + {\pounds}_{\xi_{(1,0)}} {\pounds}_{\xi_{(0,1)}} {\pounds}_{\xi_{(1,0)}}
  \right\} Q
  \nonumber\\
  && \quad\quad
  + \frac{\lambda\epsilon^{2}}{2} \left\{
    {\pounds}_{\xi_{(1,2)}}
    + 2 {\pounds}_{\xi_{(0,1)}} {\pounds}_{\xi_{(1,1)}} 
    + {\pounds}_{\xi_{(1,0)}} {\pounds}_{\xi_{(0,2)}}
    + {\pounds}_{\xi_{(0,1)}} {\pounds}_{\xi_{(1,0)}} {\pounds}_{\xi_{(0,1)}}
  \right\} Q
  \nonumber\\
  && \quad\quad
  + \frac{\epsilon^{3}}{6} \left\{
    {\pounds}_{\xi_{(0,3)}} + 3 {\pounds}_{\xi_{(0,1)}} {\pounds}_{\xi_{(0,2)}}
    + {\pounds}_{\xi_{(0,1)}}^{3}
  \right\} Q
  \nonumber\\
  && \quad
  + O^{4}(\lambda,\epsilon).
  \label{eq:two-parameter-Bruni-30-simpler}
\end{eqnarray}
Here, we note that the definitions given in
Eqs.~(\ref{eq:Bruni-14})--(\ref{eq:Bruni-22}) of the derivative
operators ${\cal L}_{(p,q)}$ and the expression 
(\ref{eq:two-parameter-Bruni-30-simpler}) of the Taylor expansion does
not include arbitrary parameters, while that derived in BGS2003 does. 
In Appendix \ref{sec:taylor-derivation}, it is shown that the
parameters in the representation derived in BGS2003 can be
eliminated through the replacement of the generators
$\xi_{(p,q)}^{a}$ without loss of generality.
The expressions (\ref{eq:Bruni-14})--(\ref{eq:Bruni-22}) and 
(\ref{eq:two-parameter-Bruni-30-simpler}) are equivalent to
those in BGS2003.
As emphasized in BGS2003, the representation of the Taylor expansion
is not unique, and there are several different but equivalent
representations.
These representations are reduced to
Eq.~(\ref{eq:two-parameter-Bruni-30-simpler}) through the 
replacement of the generators $\xi_{(p,q)}^{a}$.
Henceforth, we refer to the simpler representation
(\ref{eq:two-parameter-Bruni-30-simpler}) as the ``{\it
  canonical representation}'' of the Taylor expansion of the
two-parameter diffeomorphisms. 
Further, we denote this expression by
$\Phi^{*}_{\lambda,\epsilon}(\xi_{(p,q)}^{a})Q$ when there is a
need to specify the generators $\xi_{(p,q)}^{a}$, explicitly.

%************************************************

Next, we consider the problem how to recover the one-parameter case
from the two-parameter case when the two parameters $\lambda$ and
$\epsilon$ are no longer independent, e.g. when
$\epsilon=\epsilon(\lambda)$. 
The case in which either $\lambda$ or $\epsilon$ vanishes is
trivial and it can be recovered from the above expressions by
simply setting $\lambda=0$ or $\epsilon=0$.
Another simple case in which the two parameters $\lambda$ and $\epsilon$
are linearly dependent, i.e., $\epsilon=a\lambda$ ($a\neq 0$),
is discussed in BGS2003, there it is shown that the Taylor
expansion of the two-parameter case is reduced to the
one-parameter case through the replacement of the generators
$\xi_{(p,q)}^{a}$. 
Here, we consider the more generic case in which the infinitesimal
parameter $\epsilon$ is given by a Taylor expansion in $\lambda$ :
\begin{equation}
  \label{eq:epsilon-lambda-parameter}
  \epsilon=\epsilon(\lambda)=\sum_{n=0}^{\infty}a_{n}\frac{\lambda^{n}}{n!}.
\end{equation}
Because we only consider the representation
$\Phi^{*}_{\lambda,\epsilon}(\xi_{(p,q)}^{a})Q$ to third order,
we can restrict our attention to the expression
(\ref{eq:epsilon-lambda-parameter}) up to third order. 
Substituting (\ref{eq:epsilon-lambda-parameter}) into the Taylor
expansion of $\Phi^{*}_{\lambda,\epsilon}Q$, we obtain
\begin{eqnarray}
  \Phi^{*}_{\lambda,\epsilon}Q 
  &=& Q
  + a_{0} {\pounds}_{\xi_{(0,1)}} Q
  + \lambda {\cal L}_{\zeta_{(1,0)}} Q
  \nonumber\\
  && \quad
  + \frac{1}{2} a_{0}^{2}
  \left\{
    {\pounds}_{\xi_{(0,2)}} + {\pounds}_{\xi_{(0,1)}}^{2}
  \right\} Q
  \nonumber\\
  && \quad\quad
  + a_{0} \lambda
  \left\{ 
    {\cal L}_{\zeta_{(1,1)}} 
    + \frac{1}{2} {\cal L}_{\zeta_{(1,0)}}{\pounds}_{\xi_{(0,1)}}
    + \frac{1}{2} {\pounds}_{\xi_{(0,1)}}{\cal L}_{\zeta_{(1,0)}}
  \right\} Q
  \nonumber\\
  && \quad\quad
  + \frac{1}{2}\lambda^{2}
  \left\{
      {\cal L}_{\zeta_{(2,0)}}
     + {\cal L}_{\zeta_{(1,0)}}^{2}
  \right\} Q
  \nonumber\\
  && \quad
  + \frac{1}{6} a_{0}^{3}
  \left\{
    {\pounds}_{\xi_{(0,3)}} 
    + 3 {\pounds}_{\xi_{(0,1)}} {\pounds}_{\xi_{(0,2)}}
    + {\pounds}_{\xi_{(0,1)}}^{3}
  \right\} Q
  \nonumber\\
  && \quad\quad
  + \frac{1}{2} a_{0}^{2} \lambda
  \left\{
      {\cal L}_{\zeta_{(1,2)}}
    +2{\pounds}_{\xi_{(0,1)}}
      {\cal L}_{\zeta_{(1,1)}}
    + {\cal L}_{\zeta_{(1,0)}}
      {\pounds}_{\xi_{(0,2)}}
    + {\pounds}_{\xi_{(0,1)}}
      {\cal L}_{\zeta_{(1,0)}}
      {\pounds}_{\xi_{(0,1)}}
  \right\} Q
  \nonumber\\
  && \quad\quad
  + \frac{1}{2} a_{0} \lambda^{2}
  \left\{
    {\cal L}_{\zeta_{(2,1)}}
    +2{\cal L}_{\zeta_{(1,0)}}
      {\cal L}_{\zeta_{(1,1)}}
    + {\pounds}_{\xi_{(0,1)}}
      {\cal L}_{\zeta_{(2,0)}}
    + {\cal L}_{\zeta_{(1,0)}}
      {\pounds}_{\xi_{(0,1)}}
      {\cal L}_{\zeta_{(1,0)}}
  \right\} Q
  \nonumber\\
  && \quad\quad
  + \frac{1}{6} \lambda^{3}
  \left\{
    {\cal L}_{\zeta_{(3,0)}}
    + 3 {\cal L}_{\zeta_{(1,0)}}
        {\cal L}_{\zeta_{(2,0)}}
    + {\cal L}_{\zeta_{(1,0)}}^{3}
  \right\} Q
  \nonumber\\
  && \quad
  + O^{4}(\lambda,a_{0}),
  \label{eq:two-parameter-Bruni-30-reduced-to-one-parameter} 
\end{eqnarray}
where the vector fields $\zeta_{(p,q)}^{a}$ are defined by 
\begin{eqnarray}
  \zeta_{(1,0)}^{a} &:=& \xi_{(1,0)}^{a} + a_{1}\xi_{(0,1)}^{a}, \\
  \zeta_{(1,1)}^{a} &:=& \xi_{(1,1)}^{a} + a_{1}\xi_{(0,2)}^{a}, \\
  \zeta_{(2,0)}^{a} &:=& \xi_{(2,0)}^{a} + 2a_{1}\xi_{(1,1)}^{a} 
  + a_{2}\xi_{(0,1)}^{a} + a_{1}^{2}\xi_{(0,2)}^{a}, \\
  \zeta_{(1,2)}^{a} &:=& \xi_{(1,2)}^{a} + a_{1}\xi_{(0,3)}^{a}, \\
  \zeta_{(2,1)}^{a} &:=& \xi_{(2,1)}^{a} + 2a_{1}\xi_{(1,2)}^{a} 
  + a_{2}\xi_{(0,2)}^{a} + a_{1}^{2}\xi_{(0,3)}^{a} 
  + a_{1}[[\xi_{(0,1)},\xi_{(1,0)}],\xi_{(0,1)}]^{a}, \\
  \zeta_{(3,0)}^{a} &:=&  \xi_{(3,0)}^{a} + a_{3}\xi_{(0,1)}^{a}
  + 3a_{2}\xi_{(1,1)}^{a} + 3a_{1}\xi_{(2,1)}^{a} 
  + 3a_{1}^{2}\xi_{(1,2)}^{a}
  \nonumber\\
  && \quad\quad
  + 3a_{2}a_{1}\xi_{(0,2)}^{a} + a_{1}^{3}\xi_{(0,3)}^{a}
  + \frac{3}{2} a_{2} [\xi_{(0,1)},\xi_{(1,0)}]^{a}
  \nonumber\\
  && \quad\quad
  + 2a_{1}^{2}[[\xi_{(0,1)},\xi_{(1,0)}],\xi_{(0,1)}]^{a}
  + a_{1}[[\xi_{(1,0)},\xi_{(0,1)}],\xi_{(1,0)}]^{a}.
\end{eqnarray}
Equation
(\ref{eq:two-parameter-Bruni-30-reduced-to-one-parameter}) has
the form of a Taylor expansion of a diffeomorphism with two 
infinitesimal parameters, $\lambda$ and $a_{0}$.
Equation (\ref{eq:two-parameter-Bruni-30-reduced-to-one-parameter})
shows that the coefficient $a_{0}$ in the expansion 
(\ref{eq:epsilon-lambda-parameter}) plays the role of an
infinitesimal perturbation parameter that is independent of
$\lambda$. 
When $a_{0}=0$,
Eq.~(\ref{eq:two-parameter-Bruni-30-reduced-to-one-parameter})
reduces to the Taylor expansion in the case of a single
infinitesimal parameter $\lambda$.
Thus, even when two infinitesimal parameters depend on each
other as in the relation (\ref{eq:epsilon-lambda-parameter}), we
find that the Taylor expansion
(\ref{eq:two-parameter-Bruni-30-simpler}) is reduced to that of
a single parameter, as in the trivial case 
$\epsilon=0$ or $\lambda=0$.

%************************************************

We next derive the representation of the inverse of the
canonical representation
$\Phi^{*}_{\lambda,\epsilon}(\xi_{(p,q)}^{a})Q$ for an arbitrary
tensor field $Q$. 
To do this, we first consider the product
$\Psi_{\lambda,\epsilon}\circ\Phi_{\lambda,\epsilon}$ of the two
diffeomorphisms $\Psi_{\lambda,\epsilon}$ and
$\Phi_{\lambda,\epsilon}$.
We consider the canonical representations 
$\Psi^{*}_{\lambda,\epsilon}(\zeta_{(p,q)}^{a})Q$ and 
$\Phi^{*}_{\lambda,\epsilon}(\xi_{(p,q)}^{a})Q$.
To obtain the Taylor expansion of the pull-back of
$\Psi_{\lambda,\epsilon}\circ\Phi_{\lambda,\epsilon}$, we first
derive $\Phi^{*}_{\lambda,\epsilon}(\xi_{(p,q)}^{a})S$ for an
arbitrary tensor field $S$ and substitute the canonical
representation of the Taylor expansion
$S=\Psi^{*}_{\lambda,\epsilon}(\zeta_{(p,q)}^{a})Q$. 
Then, we obtain the representation of the pull-back,
\begin{equation}
  \left(
    \Psi_{\lambda,\epsilon}(\zeta_{(p,q)}^{a})\circ\Phi_{\lambda,\epsilon}(\xi_{(p,q)}^{a})
  \right)^{*}Q
  = 
  \Phi_{\lambda,\epsilon}^{*}(\xi_{(p,q)}^{a})\circ\Psi_{\lambda,\epsilon}^{*}(\zeta_{(p,q)}^{a})Q.
\end{equation}
In order that we can regard 
$\Psi_{\lambda,\epsilon}^{*}(\zeta_{(p,q)}^{a})$ as the 
inverse of $\Phi_{\lambda,\epsilon}^{*}(\xi_{(p,q)}^{a})$, we
choose the generators $\zeta_{(p,q)}^{a}$ so that 
$\left(
\Psi_{\lambda,\epsilon}(\zeta_{(p,q)}^{a})\circ\Phi_{\lambda,\epsilon}(\xi_{(p,q)}^{a})\right)^{*}Q=Q$.
This is accomplished by choosing
\begin{eqnarray}
  \zeta_{(1,0)}^{a} &=& - \xi_{(1,0)}^{a},  \\
  \zeta_{(0,1)}^{a} &=& - \xi_{(0,1)}^{a},  \\
  \zeta_{(2,0)}^{a} &=& - \xi_{(2,0)}^{a},  \\
  \zeta_{(0,2)}^{a} &=& - \xi_{(0,2)}^{a},  \\
  \zeta_{(1,1)}^{a} &=& - \xi_{(1,1)}^{a},  \\
  \zeta_{(3,0)}^{a} &=& - \xi_{(3,0)}^{a} + 3 [\xi_{(2,0)},\xi_{(1,0)}]^{a}, \\
  \zeta_{(2,1)}^{a} &=& - \xi_{(2,1)}^{a} + [\xi_{(2,0)},\xi_{(0,1)}]^{a} 
  + 2 [\xi_{(1,1)},\xi_{(1,0)}]^{a},  \\
  \zeta_{(1,2)}^{a} &=& - \xi_{(1,2)}^{a} + [\xi_{(0,2)},\xi_{(1,0)}]^{a}
  + 2 [\xi_{(1,1)},\xi_{(0,1)}]^{a},  \\
  \zeta_{(0,3)}^{a} &=& - \xi_{(0,3)}^{a} + 3 [\xi_{(0,2)},\xi_{(0,1)}]^{a}.
\end{eqnarray}
Then, the explicit form of 
$(\Phi_{\lambda,\epsilon}^{-1}(\xi_{(p,q)}^{a}))^{*}Q_{\lambda,\epsilon}$
is given by
\begin{eqnarray}
  && (\Phi_{\lambda,\epsilon}^{-1}(\xi_{(p,q)}^{a}))^{*}Q_{\lambda,\epsilon} 
  \nonumber\\
  &=& Q
  - \lambda {\pounds}_{\xi_{(1,0)}}Q
  - \epsilon {\pounds}_{\xi_{(0,1)}}Q
  \nonumber\\
  &&
  + \frac{\lambda^{2}}{2} \left\{
    - {\pounds}_{\xi_{(2,0)}} + {\pounds}_{\xi_{(1,0)}}^{2}
  \right\} Q_{\lambda,\epsilon}
  + \frac{\epsilon^{2}}{2} \left\{
    - {\pounds}_{\xi_{(0,2)}} + {\pounds}_{\xi_{(0,1)}}^{2}
  \right\} Q
  \nonumber\\
  &&
  + \lambda\epsilon \left\{ 
    - {\pounds}_{\xi_{(1,1)}} 
    + \frac{1}{2} {\pounds}_{\xi_{(1,0)}} {\pounds}_{\xi_{(0,1)}} 
    + \frac{1}{2} {\pounds}_{\xi_{(0,1)}} {\pounds}_{\xi_{(1,0)}} 
  \right\} Q
  \nonumber\\
  &&
  + \frac{\lambda^{3}}{6} \left\{
    - {\pounds}_{\xi_{(3,0)}} 
    + 3 {\pounds}_{\xi_{(2,0)}} {\pounds}_{\xi_{(1,0)}}
    - {\pounds}_{\xi_{(1,0)}}^{3}
  \right\} Q
  \nonumber\\
  &&
  + \frac{\lambda^{2}\epsilon}{2} \left\{
    - {\pounds}_{\xi_{(2,1)}}
    + {\pounds}_{\xi_{(2,0)}} {\pounds}_{\xi_{(0,1)}}
    + 2 {\pounds}_{\xi_{(1,1)}} {\pounds}_{\xi_{(1,0)}}
    - {\pounds}_{\xi_{(1,0)}} {\pounds}_{\xi_{(0,1)}}{\pounds}_{\xi_{(1,0)}}
  \right\} Q
  \nonumber\\
  &&
  + \frac{\lambda\epsilon^{2}}{2} \left\{
    - {\pounds}_{\xi_{(1,2)}} 
    + {\pounds}_{\xi_{(0,2)}} {\pounds}_{\xi_{(1,0)}}
    + 2 {\pounds}_{\xi_{(1,1)}} {\pounds}_{\xi_{(0,1)}} 
    - {\pounds}_{\xi_{(0,1)}} {\pounds}_{\xi_{(1,0)}}{\pounds}_{\xi_{(0,1)}}
  \right\} Q
  \nonumber\\
  &&
  + \frac{\epsilon^{3}}{6} \left\{
    - {\pounds}_{\xi_{(0,3)}} 
    + 3 {\pounds}_{\xi_{(0,2)}} {\pounds}_{\xi_{(0,1)}}
    - {\pounds}_{\xi_{(0,1)}}^{3}
  \right\} Q
  + O^{4}(\lambda,\epsilon).
  \label{eq:upto-third-ordered-gauge-transformation-X-Y-simpler}
\end{eqnarray}
This explicitly shows that
$\Phi_{\lambda,\epsilon}^{-1}\neq\Phi_{-\lambda,-\epsilon}$, as
emphasized in BGS2003.
Further, if all generators $\xi_{(p,q)}^{a}$ commute, we obtain
the equality 
$(\Phi_{\lambda,\epsilon}^{-1})^{*}(\xi_{(p,q)}^{a})Q 
= \Phi_{\lambda,\epsilon}^{*}(-\xi_{(p,q)}^{a})Q$.

%************************************************

Finally, we show that the two-parameter group of diffeomorphisms
that satisfy the property 
\begin{equation}
  \label{eq:group-property}
  \phi_{\lambda_{1},\epsilon_{1}}\circ\phi_{\lambda_{2},\epsilon_{2}}
  = \phi_{\lambda_{1}+\lambda_{2},\epsilon_{1}+\epsilon_{2}} \quad
  \forall \lambda,\epsilon\in\MR
\end{equation}
is obtained as the special case of the two-parameter family of
diffeomorphisms.
This property implies that the two-parameter group
$\phi_{\lambda,\epsilon}$ can be decomposed into two one-parameter
groups $\phi_{\lambda,0}$ and $\phi_{0,\epsilon}$ of diffeomorphisms: 
\begin{equation}
  \label{eq:group-decomposition}
  \phi_{\lambda,\epsilon} = \phi_{\lambda,0}\circ\phi_{0,\epsilon} =
  \phi_{0,\epsilon}\circ\phi_{\lambda,0}.  
\end{equation}
These two one-parameter groups of diffeomorphisms are generated
by the vector fields $\eta_{(\lambda)}^{a}$ and
$\eta_{(\epsilon)}^{a}$, respectively.
Each of these vector fields is defined by the action of the
corresponding pull-back, $\phi^{*}_{\lambda,0}$ and
$\phi^{*}_{0,\epsilon}$, for a generic tensor field $Q$ on
${\cal M}\times\MR$: 
\begin{equation}
  {\pounds}_{\eta_{(\lambda)}}Q := \lim_{\lambda\rightarrow 0}
  \frac{1}{\lambda} (\phi^{*}_{\lambda,0}Q-Q), \quad 
  {\pounds}_{\eta_{(\epsilon)}}Q := \lim_{\epsilon\rightarrow 0}
  \frac{1}{\epsilon} (\phi^{*}_{0,\epsilon}Q-Q).
\end{equation}
Because the property (\ref{eq:group-property}) implies that the
two-parameter group $\phi_{\lambda,\epsilon}$ is Abelian, the
vector field $\eta_{(\lambda)}^{a}$ and $\eta_{(\epsilon)}^{a}$
commute
\begin{equation}
  [\eta_{(\lambda)},\eta_{(\epsilon)}]^{a} = 0.
\end{equation}
The Taylor expansions of the pull-backs $\phi_{\lambda,0}^{*}T$,
$\phi_{0,\epsilon}^{*}Q$ are given
by\cite{M.Bruni-S.Soonego-CQG1997} 
\begin{eqnarray}
  \phi^{*}_{\lambda,0} Q &=& 
  \sum_{k=0}^{\infty}\frac{\lambda^{k}}{k!}
  \left[\frac{d^{k}}{d\lambda^{k}}\phi^{*}_{\lambda,0}Q\right]_{\lambda=0}
  = \sum^{\infty}_{k=0} \frac{\lambda^{k}}{k!} 
  {\pounds}_{\eta_{(\lambda)}}^{k} Q, \nonumber\\ 
  \phi^{*}_{0,\epsilon} Q &=& 
  \sum_{k=0}^{\infty}\frac{\epsilon^{k}}{k!}
  \left[\frac{d^{k}}{d\epsilon^{k}}\phi^{*}_{0,\epsilon}Q\right]_{\epsilon=0}
  = \sum^{\infty}_{k=0} \frac{\epsilon^{k}}{k!} 
  {\pounds}_{\eta_{(\epsilon)}}^{k} Q.
\end{eqnarray}
Then, using the decomposition (\ref{eq:group-decomposition}), we
obtain the Taylor expansion of the two-parameter group of
pull-backs $\phi_{\lambda,\epsilon}^{*}Q$:
\begin{equation}
  \phi_{\lambda,\epsilon}Q = \sum_{k,k'=0}^{\infty}
  \frac{\lambda^{k}\epsilon^{k'}}{k!k'!} \left[
  \frac{\partial^{k+k'}}{\partial\lambda^{k}\partial\epsilon^{k'}}\phi^{*}_{\lambda,\epsilon}Q
  \right]_{\lambda=\epsilon=0} 
  = \sum^{\infty}_{k,k'=0}\frac{\lambda^{k}\epsilon^{k'}}{k!k'!}
  {\pounds}^{k}_{\eta_{(\lambda)}}{\pounds}^{k'}_{\eta_{(\epsilon)}}Q.
\end{equation}
This expression is also obtained as the special case of the Taylor
expansion (\ref{eq:two-parameter-Bruni-30-simpler}) of the
two-parameter family of diffeomorphisms 
$\Phi_{\lambda,\epsilon}^{*}Q$ imposing the conditions that 
$\eta_{(\lambda)}^{a} = \xi_{(1,0)}^{a}\neq 0$, 
$\eta_{(\epsilon)}^{a} = \xi_{(0,1)}^{a}\neq 0$, 
$[\xi_{(1,0)},\xi_{(0,1)}]^{a}=0$, and $\xi_{(p,q)}^{a}=0$ for
$p+q>1$.

%%%%%%%%%%%%%%%%%%%%%%%%%%%%%%%%%%%%%%%%%%%%%%%%%%%%%%%%%%%%%%%%%%%%%
\section{Gauge transformation of perturbation variables}
\label{sec:gauge-trans}
%%%%%%%%%%%%%%%%%%%%%%%%%%%%%%%%%%%%%%%%%%%%%%%%%%%%%%%%%%%%%%%%%%%%%

%************************************************

Using the above Taylor expansion of two-parameter diffeomorphisms, we
consider gauge transformations in two-parameter perturbations of
the manifold.

%************************************************

%%%%%%%%%%%%%%%%%%%%%%%%%%%%%%%%%%%%%%%%%%%%%%%%%%%%%%%%%%%%%%%%%%%%%
\subsection{Gauges in perturbation theory}
\label{sec:gauges}
%%%%%%%%%%%%%%%%%%%%%%%%%%%%%%%%%%%%%%%%%%%%%%%%%%%%%%%%%%%%%%%%%%%%%

%************************************************

Let us consider the spacetime $({\cal M}_{0},{}^{(0)}g_{ab})$, which
is the background spacetime for the perturbations, and a physical
spacetime $({\cal M},g_{ab})$, which we attempt to describe as a 
perturbation of the background spacetime 
$({\cal M}_{0},{}^{(0)}{g}_{ab})$.  
Let us formally denote the spacetime metric and the other physical
tensor fields on the physical spacetime ${\cal M}$ by $Q$.
In perturbation theory, we are used to write expressions of the
form 
\begin{equation}
  \label{eq:variable-symbolic-perturbation}
  Q(x) = Q_{0}(x) + \delta Q(x).
\end{equation}
This expression relates the variable $Q$ on ${\cal M}$
to the background value of the same field, $Q_{0}$, and the
perturbation $\delta Q$. 
In the expression (\ref{eq:variable-symbolic-perturbation}), we
have implicitly assigned a correspondence between points of the
perturbed and the background spacetime.
This is the implicit assumption of the existence of a map 
${\cal M}_{0}\rightarrow{\cal M}$ $:$  
$p\in{\cal M}_{0}\mapsto q\in{\cal M}$\cite{J.M.Stewart-M.Walker11974}.
This correspondence associated with the map 
${\cal M}_{0}\rightarrow{\cal M}$ is what is usually called a
``gauge choice'' in the context of perturbation
theory.\footnote{
  More precisely, as mentioned in BGS2003,
  Eq.~(\ref{eq:variable-symbolic-perturbation}) gives a relation
  between the images of the fields in $\MR^{m}$ rather than between
  the fields themselves on the respective manifolds ${\cal M}$ and 
  ${\cal M}_{0}$, i.e., we are saying that there is a unique point
  $x\in\MR$ that is at the same time the image of two points: one in
  ${\cal M}_{0}$ and another in ${\cal M}$. 
  However, in this paper, we deal with only the point identification
  map ${\cal X}:{\cal M}_{0}\rightarrow{\cal M}$.
  }
Clearly, this is more than the usual assignment of coordinate
labels to points on the single spacetime. 
It is important to note that the correspondence established by
such a relation as Eq.~(\ref{eq:variable-symbolic-perturbation})
is not unique.
Rather, Eq.~(\ref{eq:variable-symbolic-perturbation})
involves the degree of freedom corresponding to the choice of
the map ${\cal M}_{0}\rightarrow{\cal M}$ (the choice of the
point identification map ${\cal M}_{0}\rightarrow{\cal M}$). 
This is called ``gauge freedom''.
Further, such freedom always exists in the perturbation of a theory
in which we impose general covariance.

%************************************************

Here, we introduce an $(m+2)$-dimensional manifold ${\cal N}$ to
study two-parameter perturbation theory based on the above idea.  
The manifold ${\cal N}$ is foliated into $m$-dimensional
submanifolds diffeomorphic to ${\cal M}$, so that 
${\cal N}={\cal M}\times\MR^{2}$.  
Each copy of ${\cal M}$ is labeled by the corresponding value of
the parameters $(\lambda,\epsilon)\in\MR^{2}$. 
The manifold ${\cal N}$ has a natural differentiable
structure that is the direct product of that of ${\cal M}$
and $\MR^{2}$.
With this construction, the perturbed spacetimes 
${\cal M}_{\lambda,\epsilon}$ for each $(\lambda,\epsilon)$ must
have the same differential structure, and the changes of the
differential structure resulting from the perturbation, for
example the formation of singularities, is excluded from our
consideration. 
Each point on ${\cal N}$ is assigned by $(p,\lambda,\epsilon)$,
where $p\in{\cal M}_{\lambda,\epsilon}$, and each point on the
background spacetime ${\cal M}_{0}$ in ${\cal N}$ is assigned by 
$\lambda=\epsilon=0$.

%************************************************

Let us consider the set of field equations 
\begin{equation}
  \label{eq:field-eq-for-Q}
  {\cal E}[Q_{\lambda,\epsilon}] = 0
\end{equation}
on ${\cal M}_{\lambda,\epsilon}$ for the physical variables $Q$ on
${\cal M}_{\lambda,\epsilon}$. 
The field equation (\ref{eq:field-eq-for-Q}) formally represents
the Einstein equation for the metric on ${\cal M}_{\lambda,\epsilon}$
and the field equations for matter fields on 
${\cal M}_{\lambda,\epsilon}$.
If a tensor field $Q_{\lambda,\epsilon}$ is given on each 
${\cal M}_{\lambda,\epsilon}$, $Q_{\lambda,\epsilon}$ is
automatically extended to a tensor field on ${\cal N}$ by 
$Q(p,\lambda,\epsilon) := Q_{\lambda,\epsilon}(p)$, with 
$p\in{\cal M}_{\lambda,\epsilon}$. 
In this extension, the field equation (\ref{eq:field-eq-for-Q}) is
regarded as the equation on ${\cal N}$.

%************************************************

Now, we define the perturbation for an arbitrary tensor field
$Q$ by comparing $Q_{\lambda,\epsilon}$ with $Q_{0}$.
To do this, it is necessary to identify the points of
${\cal M}_{\lambda,\epsilon}$ with those of ${\cal M}_{0}$.
This is easily accomplished by assigning a diffeomorphism 
${\cal X}_{\lambda,\epsilon}:{\cal N}\rightarrow{\cal N}$ such that
${\cal X}_{\lambda,\epsilon} : 
{\cal M}_{0}\rightarrow{\cal M}_{\lambda,\epsilon}$. 
It is natural to regard ${\cal X}_{\lambda,\epsilon}$ as
one of the two-parameter group of diffeomorphisms that satisfy
the property (\ref{eq:group-property}).
Then, ${\cal X}_{\lambda,\epsilon}$ is generated by two vector
fields ${}^{\cal X}\eta_{(\lambda)}^{a}$ and 
${}^{\cal X}\eta_{(\epsilon)}^{a}$ on ${\cal N}$ that satisfy  
\begin{equation}
  \label{eq:commute-each-other}
  [{}^{\cal X}\eta_{(\lambda)},{}^{\cal X}\eta_{(\epsilon)}]^{a} = 0.
\end{equation}
Further, the normal forms of ${\cal M}_{\lambda,\epsilon}$ in
${\cal N}$ are given by $(d\lambda)_{a}$ and $(d\epsilon)_{a}$,
and their duals are defined by 
\begin{equation}
  \left(\frac{\partial}{\partial\lambda}\right)^{a}(d\lambda)_{a}
  = 1, \quad
  \left(\frac{\partial}{\partial\epsilon}\right)^{a}(d\epsilon)_{a}
  = 1, \quad
  \left(\frac{\partial}{\partial\lambda}\right)^{a}(d\epsilon)_{a}
  = 0, \quad
  \left(\frac{\partial}{\partial\epsilon}\right)^{a}(d\lambda)_{a}
  = 0.
\end{equation}
The vector fields ${}^{\cal X}\eta_{(\lambda)}^{a}$ and 
${}^{\cal X}\eta_{(\epsilon)}^{a}$ are chosen so that 
\begin{equation}
  \label{eq:generator-decomp-def}
  {}^{\cal X}\eta_{(\lambda)}^{a}
  = \left(\frac{\partial}{\partial\lambda}\right)^{a} +
  \theta_{(\lambda)}^{a}, \quad 
  {}^{\cal X}\eta_{(\epsilon)}^{a} 
  = \left(\frac{\partial}{\partial\epsilon}\right)^{a} +
  \theta_{(\epsilon)}^{a}, 
\end{equation}
where $\theta^{a}_{(\lambda)}$ and $\theta^{a}_{(\epsilon)}$ are
tangent to ${\cal M}_{\lambda,\epsilon}$ for each $\lambda$ and
$\epsilon$ :
\begin{eqnarray}
  \label{eq:orthogonality-of-theta}
  \theta_{(\lambda),(\epsilon)}^{a}(d\epsilon)_{a} 
  = \theta_{(\lambda),(\epsilon)}^{a}(d\lambda)_{a} = 0. 
\end{eqnarray}
The choice of the vector fields
$\theta_{(\lambda),(\epsilon)}^{a}$ is essentially arbitrary,
except for the conditions (\ref{eq:commute-each-other}) and
(\ref{eq:orthogonality-of-theta}). 
Therefore, we choose $\theta_{(\lambda),(\epsilon)}^{a}$ so that 
\begin{equation}
  {\pounds}_{\frac{\partial}{\partial\lambda}}
  \theta_{(\lambda),(\epsilon)}^{a}
  = 
  {\pounds}_{\frac{\partial}{\partial\epsilon}}
  \theta_{(\epsilon),(\lambda)}^{a}
  = 0
\end{equation}
for simplicity.
Except for these conditions, we can regard
$\theta_{(\lambda),(\epsilon)}^{a}$ as essentially arbitrary
vector fields on ${\cal M}_{\lambda,\epsilon}$ (not on ${\cal N}$) 
that all commute.

%************************************************

The perturbation $\Delta^{\cal X}_{0}Q_{\lambda,\epsilon}$ of a tensor
field $Q$ for a gauge choice ${\cal X}$ can now be defined simply as
\begin{equation}
  \label{eq:Bruni-34}
  \Delta^{\cal X}_{0}Q_{\lambda,\epsilon} :=
  \left.{\cal X}^{*}_{\lambda,\epsilon}Q\right|_{{\cal M}_{0}} - Q_{0}.
\end{equation}
The first term on the right-hand side of
(\ref{eq:Bruni-34}) can be Taylor-expanded as
\begin{equation}
  \label{eq:Bruni-35}
  \left.{\cal X}^{*}_{\lambda,\epsilon}Q\right|_{{\cal M}_{0}}
  =
  \sum^{\infty}_{k,k'=0} \frac{\lambda^{k}\epsilon^{k'}}{k!k'!}
  \delta^{(k,k')}_{\cal X}Q,
\end{equation}
where
\begin{equation}
  \label{eq:Bruni-36}
  \delta^{(k,k')}_{\cal X}Q := 
  \left[
    \frac{\partial^{k+k'}}{\partial\lambda^{k}\partial\epsilon^{k'}}
    {\cal X}^{*}_{\lambda,\epsilon}Q
  \right]_{\lambda=\epsilon=0} = 
  \left.
    {\cal L}_{{}^{\cal X}\eta_{\lambda}}^{k} 
    {\cal L}_{{}^{\cal X}\eta_{\epsilon}}^{k'} Q
  \right|_{{\cal M}_{0}}.
\end{equation}
Equations (\ref{eq:Bruni-34})--(\ref{eq:Bruni-36}) define the
perturbation of order $(k,k')$ of a physical variable $Q$ for
the gauge choice ${\cal X}$ and its background value 
$\delta^{(0,0)}_{\cal X}Q=Q_{0}$.

%************************************************

%%%%%%%%%%%%%%%%%%%%%%%%%%%%%%%%%%%%%%%%%%%%%%%%%%%%%%%%%%%%%%%%%%%%%
\subsection{Gauge invariance and gauge transformations}
\label{sec:Bruni-gauge-inv-gauge-trans}
%%%%%%%%%%%%%%%%%%%%%%%%%%%%%%%%%%%%%%%%%%%%%%%%%%%%%%%%%%%%%%%%%%%%%

%************************************************

\begin{figure}
  \begin{center}
    \includegraphics[width=0.8\textwidth]{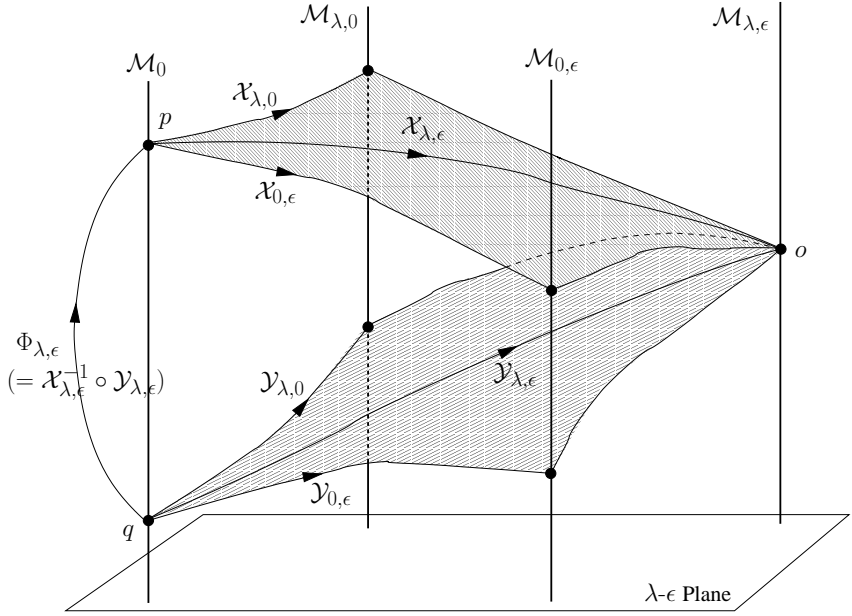}
  \end{center}
  \caption{Point identification from the background manifold to the
    perturbed manifold.} 
  \label{fig:figure1}
\end{figure}

Let us now suppose that two gauges ${\cal X}$ and ${\cal Y}$ are
generated by the pairs of vector fields
$({}^{\cal X}\eta_{(\lambda)},{}^{\cal X}\eta_{(\epsilon)})$ and 
$({}^{\cal Y}\eta_{(\lambda)},{}^{\cal Y}\eta_{(\epsilon)})$,
respectively.
These vector field are defined on ${\cal N}$ as the vector
fields in Eqs.~(\ref{eq:generator-decomp-def}) :
\begin{eqnarray}
  && {}^{\cal X}\eta_{(\lambda)}^{a}
  = \left(\frac{\partial}{\partial\lambda}\right)^{a} +
  \theta_{(\lambda)}^{a}, \quad 
  {}^{\cal X}\eta_{(\epsilon)}^{a} 
  = \left(\frac{\partial}{\partial\epsilon}\right)^{a} +
  \theta_{(\epsilon)}^{a}, 
  \\
  && {}^{\cal Y}\eta_{(\lambda)}^{a}
  = \left(\frac{\partial}{\partial\lambda}\right)^{a} +
  \iota_{(\lambda)}^{a}, \quad 
  {}^{\cal Y}\eta_{(\epsilon)}^{a} 
  = \left(\frac{\partial}{\partial\epsilon}\right)^{a} +
  \iota_{(\epsilon)}^{a}. 
\end{eqnarray}
The condition (\ref{eq:commute-each-other}) for each set of
generators, ${}^{\cal X}\eta_{(\lambda),(\epsilon)}^{a}$ and 
${}^{\cal Y}\eta_{(\lambda),(\epsilon)}^{a}$, 
implies that the two-dimensional tangent space spanned by 
${}^{\cal X}\eta_{(\lambda),(\epsilon)}^{a}$ 
(${}^{\cal X}\eta_{(\lambda),(\epsilon)}^{a}$) possesses a
two-dimensional integral surface.
These integral surfaces of
${}^{\cal X}\eta_{(\lambda),(\epsilon)}^{a}$ and 
${}^{\cal Y}\eta_{(\lambda),(\epsilon)}^{a}$ define two
two-parameter groups of diffeomorphisms ${\cal X}$ and 
${\cal Y}$ on ${\cal N}$. 
Further, ${}^{\cal X}\eta_{(\lambda),(\epsilon)}^{a}$ and 
${}^{\cal Y}\eta_{(\lambda),(\epsilon)}^{a}$ are everywhere
transverse to ${\cal M}_{\lambda,\epsilon}$, and points lying on 
the same integral surface of either of the two are to be
regarded as {\it the same point} within the respective gauge.
(See Fig.~\ref{fig:figure1}.)
Then, ${\cal X}$ and ${\cal Y}$ are both point identification maps.
When
$\theta_{(\lambda),(\epsilon)}^{a}\neq\iota_{(\lambda),(\epsilon)}^{a}$,
these point identification maps are regarded as two different gauge
choices. 
% \begin{figure}[htbp]
%   \begin{center}
%     \leavevmode
%     \epsfxsize=0.5\textwidth
%     \epsfbox[0 0 903 602]{gauge-inv-fig1.eps}
%     \caption{Point identification from the background manifold to the
%     perturbed manifold.}
%     \label{fig:figure1}
%   \end{center}
% \end{figure}

%************************************************

The pairs of vector fields
${}^{\cal X}\eta_{(\lambda),(\epsilon)}^{a}$ and
${}^{\cal Y}\eta_{(\lambda),(\epsilon)}^{a}$ are both generators of
two-parameter groups of diffeomorphisms, and they pull back a generic 
tensor $Q$ to two other tensor fields, 
${\cal X}_{\lambda,\epsilon}^{*}Q$ and 
${\cal Y}_{\lambda,\epsilon}^{*}Q$, for any given value of
$(\lambda,\epsilon)$.
In particular, on ${\cal M}_{0}$, we now have three
tensor fields, i.e. $Q_{0}$ and the following two :
\begin{equation}
  {}^{\cal X}Q_{\lambda,\epsilon} := 
  \left.{\cal X}^{*}_{\lambda,\epsilon}Q\right|_{{\cal M}_{0}},
  \quad
  {}^{\cal Y}Q_{\lambda,\epsilon} := 
  \left.{\cal Y}^{*}_{\lambda,\epsilon}Q\right|_{{\cal M}_{0}}.
  \label{eq:Bruni-38}
\end{equation}
Because ${\cal X}$ and ${\cal Y}$ represent gauge choices mapping 
the background spacetime ${\cal M}_{0}$ into a perturbed manifold
${\cal M}_{\lambda,\epsilon}$, as mentioned above,
${}^{\cal X}Q_{\lambda,\epsilon}$ and 
${}^{\cal Y}Q_{\lambda,\epsilon}$ are the representations in 
${\cal M}_{0}$ of the perturbed tensor for the two gauges. 
Using (\ref{eq:Bruni-34})--(\ref{eq:Bruni-36}), we can write 
\begin{eqnarray}
  \label{eq:Bruni-39}
  {}^{\cal X}Q_{\lambda,\epsilon} &=& \sum^{\infty}_{k,k'=0}
  \frac{\lambda^{k}\epsilon^{k'}}{k!k'!}
  \delta^{(k,k')}_{{\cal X}}Q
  = Q_{0} + \Delta^{\cal X}_{0}Q_{\lambda,\epsilon}, \\
  \label{eq:Bruni-40}
  {}^{\cal Y}Q_{\lambda,\epsilon} &=& \sum^{\infty}_{k,k'=0}
  \frac{\lambda^{k}\epsilon^{k'}}{k!k'!}
  \delta^{(k,k')}_{\cal Y}Q
  = Q_{0} + \Delta^{\cal Y}_{0}Q_{\lambda,\epsilon},
\end{eqnarray}
where $\delta^{(k,k')}_{\cal X}Q$, and $\delta^{(k,k')}_{\cal Y}Q$
are the perturbations (\ref{eq:Bruni-36}) in the gauges
${\cal X}$ and ${\cal Y}$, respectively ;
\begin{eqnarray}
  \label{eq:Bruni-41}
  \delta^{(k,k')}_{\cal X}Q = 
  \left.
    {\pounds}_{{}^{\cal X}\eta_{(\lambda)}}^{k}
    {\pounds}_{{}^{\cal X}\eta_{(\epsilon)}}^{k'} Q
  \right|_{{\cal M}_{0}}, \quad
  \delta^{(k,k')}_{\cal Y}Q = \left.
    {\pounds}_{{}^{\cal Y}\eta_{(\lambda)}}^{k}
    {\pounds}_{{}^{\cal Y}\eta_{(\epsilon)}}^{k'} Q
  \right|_{{\cal M}_{0}}.
\end{eqnarray}

%************************************************

Following Bruni et al.\cite{Bruni-Gualtieri-Sopuerta}, we consider the
concept of {\it gauge invariance up to order ($n,n'$)}.
We say that $Q$ is {\it gauge invariant up to order ($n,n'$)}
iff for any two gauges ${\cal X}$ and ${\cal Y}$
\begin{equation}
  \delta^{(k,k')}_{\cal X}Q = \delta^{(k,k')}_{\cal Y}Q \quad 
  \forall (k,k'), \quad \mbox{with} \quad k<n, \quad k'<n'.
\end{equation}
From this definition, we can prove that the $(n,n')$-order
perturbation of a tensor field $Q$ is gauge invariant up to order
$(n,n')$ iff in a given gauge ${\cal X}$ we have ${\pounds}_{\xi}
{}^{\cal X}\delta^{(k,k')}Q = 0$ for any vector field $\xi^{a}$ defined
on ${\cal M}_{0}$ and for any
$(k,k')<(n,n')$.
As a consequence, the $(n,n')$-order perturbation of a tensor
field $Q$ is gauge invariant up to order $(n,n')$ iff $Q_{0}$
and all its perturbations of lower than $(n,n')$ order are, in
any gauge, either vanishing or constant scalars, or a
combination of Kronecker deltas with constant 
coefficients\cite{J.M.Stewart-M.Walker11974,Bruni-Gualtieri-Sopuerta,M.Bruni-S.Soonego-CQG1997}.

%************************************************

Next, we consider the gauge transformation of a tensor field $Q$.
If a tensor $Q$ is not gauge invariant, its representation on 
${\cal M}_{0}$ does change under a gauge transformation.
To consider the change under a gauge transformation, we introduce the
diffeomorphism 
$\Phi_{\lambda,\epsilon} : {\cal M}_{0} \rightarrow {\cal M}_{0}$ for
each value of $(\lambda,\epsilon)\in\MR^{2}$.
The diffeomorphism $\Phi_{\lambda,\epsilon}$ is defined by 
\begin{equation}
  \label{eq:diffeo-def-from-Xinv-Y}
  \Phi_{\lambda,\epsilon} :=
  ({\cal X}_{\lambda,\epsilon})^{-1}\circ{\cal Y}_{\lambda,\epsilon} 
  = {\cal X}_{-\lambda,-\epsilon}\circ{\cal Y}_{\lambda,\epsilon}, 
\end{equation}
where we have used the fact that the point identification map ${\cal X}$ is
a two-parameter group of diffeomorphism.
When ${\cal X}$ and ${\cal Y}$ are regarded as two different gauge
choices, $\Phi_{\lambda,\epsilon}$ represents the gauge transformation
from the gauge ${\cal X}$ to the gauge ${\cal Y}$.
It is important to note that as a consequence of the definition
(\ref{eq:diffeo-def-from-Xinv-Y}), 
$\Phi_{\lambda,\epsilon}:{\cal M}_{0}\times\MR^{2}\rightarrow{\cal M}_{0}$ 
does {\it not} become a two-parameter group of diffeomorphisms in
${\cal M}_{0}$, while the identification maps ${\cal X}$ and 
${\cal Y}$ are both two-parameter group of diffeomorphisms.
Actually, 
$\Phi_{\lambda_{1},\epsilon_{1}}\circ\Phi_{\lambda_{2},\epsilon_{2}}
\neq \Phi_{\lambda_{1}+\lambda_{2},\epsilon_{1}+\epsilon_{2}}$, 
due to the fact that the vector fields 
${}^{\cal X}\eta_{(\lambda),(\epsilon)}^{a}$ and
${}^{\cal Y}\eta_{(\lambda),(\epsilon)}^{a}$ have, in general, a
non-vanishing commutator.
In spite that $\Phi_{\lambda,\epsilon}$ is not a two-parameter
group of diffeomorphisms, the representation of the Taylor
expansion of the pull-back $\Phi_{\lambda,\epsilon}^{*}Q$ for an
arbitrary tensor field $Q$ is obtained by using the
results of \S\ref{sec:Taylor}.

%************************************************

The tensor fields ${}^{\cal X}Q_{\lambda,\epsilon}$ and
${}^{\cal Y}Q_{\lambda,\epsilon}$, which are defined on ${\cal M}_{0}$
by the gauges ${\cal X}$ and ${\cal Y}$, are connected by the linear
map $\Phi^{*}_{\lambda,\epsilon}$ as
\begin{eqnarray}
  {}^{\cal Y}Q_{\lambda,\epsilon} &=&
  \left.{\cal Y}^{*}_{\lambda,\epsilon}Q\right|_{{\cal M}_{0}} 
  =
  \left.\left(
      {\cal Y}^{*}_{\lambda,\epsilon}
      {\cal X}^{*}_{-\lambda,-\epsilon}
      {\cal X}^{*}_{\lambda,\epsilon}Q\right)
  \right|_{{\cal M}_{0}} \nonumber\\
  &=&
  \left.\Phi^{*}_{\lambda,\epsilon}\left(
      {\cal X}^{*}_{\lambda,\epsilon}Q\right) 
  \right|_{{\cal M}_{0}}
  =  \Phi^{*}_{\lambda,\epsilon} {}^{\cal X}Q_{\lambda,\epsilon}.
  \label{eq:Bruni-45} 
\end{eqnarray}
Therefore, the gauge transformation to an arbitrary order
$(n,n')$ is given by the Taylor expansion of the pull-back
$\Phi^{*}_{\lambda,\epsilon}Q$, whose terms are explicitly given
in \S\ref{sec:Taylor}.
Up to third order, the explicit form of the Taylor expansion 
is given by 
\begin{eqnarray}
  \Phi^{*}_{\lambda,\epsilon}Q &=& Q
  + \lambda {\pounds}_{\xi_{(1,0)}}Q
  + \epsilon {\pounds}_{\xi_{(0,1)}}Q
  \nonumber\\
  && \quad
  + \frac{\lambda^{2}}{2} \left\{{\pounds}_{\xi_{(2,0)}} + 
    {\pounds}_{\xi_{(1,0)}}^{2}\right\} Q
  \nonumber\\
  && \quad\quad
  + \lambda\epsilon \left\{ {\pounds}_{\xi_{(1,1)}} 
    + \frac{1}{2} {\pounds}_{\xi_{(1,0)}} {\pounds}_{\xi_{(0,1)}} 
    + \frac{1}{2} {\pounds}_{\xi_{(0,1)}} {\pounds}_{\xi_{(1,0)}} 
  \right\} Q
  \nonumber\\
  && \quad\quad
  + \frac{\epsilon^{2}}{2} \left\{{\pounds}_{\xi_{(0,2)}} + 
    {\pounds}_{\xi_{(0,1)}}^{2}\right\} Q
  \nonumber\\
  && \quad
  + \frac{\lambda^{3}}{6} \left\{
    {\pounds}_{\xi_{(3,0)}} + 3 {\pounds}_{\xi_{(1,0)}} {\pounds}_{\xi_{(2,0)}}
    + {\pounds}_{\xi_{(1,0)}}^{3}
  \right\} Q
  \nonumber\\
  && \quad\quad
  + \frac{\lambda^{2}\epsilon}{2} \left\{
    {\pounds}_{\xi_{(2,1)}}
    + 2 {\pounds}_{\xi_{(1,0)}} {\pounds}_{\xi_{(1,1)}} 
    + {\pounds}_{\xi_{(0,1)}} {\pounds}_{\xi_{(2,0)}}
    + {\pounds}_{\xi_{(1,0)}} {\pounds}_{\xi_{(0,1)}} {\pounds}_{\xi_{(1,0)}}
  \right\} Q
  \nonumber\\
  && \quad\quad
  + \frac{\lambda\epsilon^{2}}{2} \left\{
    {\pounds}_{\xi_{(1,2)}}
    + 2 {\pounds}_{\xi_{(0,1)}} {\pounds}_{\xi_{(1,1)}} 
    + {\pounds}_{\xi_{(1,0)}} {\pounds}_{\xi_{(0,2)}}
    + {\pounds}_{\xi_{(0,1)}} {\pounds}_{\xi_{(1,0)}} {\pounds}_{\xi_{(0,1)}}
  \right\} Q
  \nonumber\\
  && \quad\quad
  + \frac{\epsilon^{3}}{6} \left\{
    {\pounds}_{\xi_{(0,3)}} + 3 {\pounds}_{\xi_{(0,1)}} {\pounds}_{\xi_{(0,2)}}
    + {\pounds}_{\xi_{(0,1)}}^{3}
  \right\} Q
  \nonumber\\
  && \quad
  + O^{4}(\lambda,\epsilon),
  \label{eq:Bruni-46} 
\end{eqnarray}
using the canonical representation
(\ref{eq:two-parameter-Bruni-30-simpler}) of
$\Phi^{*}_{\lambda,\epsilon}(\xi^{a}_{(p,q)})$, where
$\xi_{(p,q)}$ are now the generators of the gauge transformation
$\Phi_{\lambda,\epsilon}$.

%************************************************

Comparing the representation (\ref{eq:Bruni-46}) of the expansion in
terms of the generators $\xi_{(p,q)}$ of the pull-back
$\Phi_{\lambda,\epsilon}^{*}Q$ and that in terms of the generators 
${}^{\cal X}\eta_{(\lambda),(\epsilon)}^{a}$ and 
${}^{\cal Y}\eta_{(\lambda),(\epsilon)}^{a}$ of the pull-back
${\cal Y}^{*}_{\lambda,\epsilon}\circ{\cal X}^{*}_{-\lambda,-\epsilon}Q$
($=\Phi_{\lambda,\epsilon}^{*}Q$), we find explicit
expressions for  the generators $\xi_{(p,q)}^{a}$ of the gauge
transformation $\Phi={\cal X}^{-1}\circ{\cal Y}$ in terms of the
gauge vector fields ${}^{\cal X}\eta_{(\lambda),(\epsilon)}^{a}$ and
${}^{\cal Y}\eta_{(\lambda),(\epsilon)}^{a}$.
Here, we give their expressions up to third order:
\begin{eqnarray}
  \xi_{(1,0)}^{a} &=& {}^{\cal Y}\eta_{(\lambda)}^{a}
  - {}^{\cal X}\eta_{(\lambda)}^{a} 
  \nonumber\\
  &=& \iota_{(\lambda)}^{a} - \theta_{(\lambda)}^{a}
  \label{eq:Bruni-61}
  , \\
  \xi_{(0,1)}^{a} &=& {}^{\cal Y}\eta_{(\epsilon)}^{a}
  - {}^{\cal X}\eta_{(\epsilon)}^{a}
  \nonumber\\
  &=& \iota_{(\epsilon)}^{a} - \theta_{(\epsilon)}^{a}
  \label{eq:Bruni-62}
  , \\
  \xi_{(2,0)}^{a} &=& \left[{}^{\cal X}\eta_{(\lambda)},
    {}^{\cal Y}\eta_{(\lambda)}\right]^{a}
  \nonumber\\
  &=& \left[\theta_{(\lambda)}, \iota_{(\lambda)}\right]^{a}
  \label{eq:Bruni-63}
  , \\
  \xi_{(1,1)}^{a} &=&   
  \frac{1}{2}
  [
   {}^{\cal X}\eta_{(\lambda)}, {}^{\cal Y}\eta_{(\epsilon)}
  ]^{a}
  +
  \frac{1}{2}
  [
  {}^{\cal X}\eta_{(\epsilon)}, {}^{\cal Y}\eta_{(\lambda)}
  ]^{a}
  \nonumber\\
  &=& \frac{1}{2}[
  \theta_{(\lambda)},\iota_{(\epsilon)}
  ]^{a}
  + 
  \frac{1}{2}[
  \theta_{(\lambda)}, \iota_{(\epsilon)}
  ]^{a}
  \label{eq:Bruni-64}
  , \\
  \xi_{(0,2)}^{a} &=& \left[{}^{\cal X}\eta_{(\epsilon)},
    {}^{\cal Y}\eta_{(\epsilon)}\right]^{a}
  \nonumber\\
  &=& \left[\theta_{(\epsilon)}, \iota_{(\epsilon)}\right]^{a}
  \label{eq:Bruni-65}
  , \\
  \xi_{(3,0)}^{a}
  &=& 
  [[{}^{\cal X}\eta_{(\lambda)},{}^{\cal Y}\eta_{(\lambda)}],
  {}^{\cal Y}\eta_{(\lambda)}-2{}^{\cal X}\eta_{(\lambda)}]^{a}
  \nonumber\\
  &=& 
  [[\theta_{(\lambda)}, \iota_{(\lambda)}],
  \iota_{(\lambda)}-2\theta_{(\lambda)}]^{a}
  \label{eq:gene-rela-3.0}
  , \\
  \xi_{(2,1)}^{a}
  &=& 
  \left[
    \left[
      {}^{\cal Y}\eta_{(\lambda)},{}^{\cal X}\eta_{(\epsilon)}
    \right],
    {}^{\cal X}\eta_{(\lambda)}
  \right]^{a}
  +
  \left[
    \left[
      {}^{\cal X}\eta_{(\lambda)},{}^{\cal Y}\eta_{(\epsilon)}
    \right],
    {}^{\cal Y}\eta_{(\lambda)}
    - {}^{\cal X}\eta_{(\lambda)}
  \right]^{a}
  \nonumber\\
  &=& 
  \left[
    \left[
      \iota_{(\lambda)},\theta_{(\epsilon)}
    \right],
    \theta_{(\lambda)}
  \right]^{a}
  +
  \left[
    \left[
      \theta_{(\lambda)},\iota_{(\epsilon)}
    \right],
    \iota_{(\lambda)} - \theta_{(\lambda)}
  \right]^{a}
  \label{eq:gene-rela-2.1}
  , \\
  \xi_{(1,2)}^{a}
  &=& 
  \left[
    \left[
      {}^{\cal Y}\eta_{(\epsilon)},{}^{\cal X}\eta_{(\epsilon)}
    \right],
    {}^{\cal X}\eta_{(\lambda)}
  \right]^{a}
  +
  \left[
    \left[
      {}^{\cal X}\eta_{(\epsilon)},{}^{\cal Y}\eta_{(\epsilon)}
    \right],
    {}^{\cal Y}\eta_{(\epsilon)} - {}^{\cal X}\eta_{(\epsilon)}
  \right]^{a}
  \nonumber\\
  &=& 
  \left[
    \left[
      \iota_{(\epsilon)},\theta_{(\epsilon)}
    \right],
    \theta_{(\lambda)}
  \right]^{a}
  +
  \left[
    \left[
      \theta_{(\epsilon)},\iota_{(\epsilon)}
    \right],
    \iota_{(\epsilon)} - \theta_{(\epsilon)}
  \right]^{a}
  \label{eq:gene-rela-1.2}
  , \\
  \xi_{(0,3)}^{a}
  &=& 
  [[{}^{\cal X}\eta_{(\epsilon)},{}^{\cal Y}\eta_{(\epsilon)}],
  {}^{\cal Y}\eta_{(\epsilon)}-2{}^{\cal X}\eta_{(\epsilon)}]^{a}
  \nonumber\\
  &=& 
  [[\theta_{(\epsilon)}, \iota_{(\epsilon)}],
  \iota_{(\epsilon)}-2\theta_{(\epsilon)}]^{a}
  \label{eq:gene-rela-0.3}
  .
\end{eqnarray}
The expression (\ref{eq:Bruni-64}) of the generator
$\xi_{(1,1)}^{a}$ is different from that derived in BGS2003.
This is due to the difference between the representations
of the Taylor expansion of the pull-back
$\Phi^{*}_{\lambda,\epsilon}Q$. 
In the perturbation theory, these expressions,
(\ref{eq:Bruni-61})--(\ref{eq:gene-rela-0.3}), are evaluated on
the background spacetime ${\cal M}_{0}$.
Then, these expressions show explicitly that the generators
$\xi_{(p,q)}^{a}$ of the gauge transformation
$\Phi_{\lambda,\epsilon}={\cal X}^{-1}\circ{\cal Y}$
are vector fields on the background ${\cal M}_{0}$.
Further, Eqs.~(\ref{eq:Bruni-63})--(\ref{eq:gene-rela-0.3}) show that
the generators $\xi_{(p,q)}^{a}$ with $p+q>1$ naturally arise from the
non-commutativity of the gauge generators 
${}^{\cal X}\eta_{(\lambda),(\epsilon)}^{a}$ and
${}^{\cal Y}\eta_{(\lambda),(\epsilon)}^{a}$.

%************************************************

We can now derive the relation between the perturbations in the two
different gauges.
Up to order $(n,n')$ with $n+n'\leq 3$, these relations can be
derived by substituting (\ref{eq:Bruni-39}) and
(\ref{eq:Bruni-40}) into (\ref{eq:Bruni-46}):
\begin{eqnarray}
  \label{eq:Bruni-47} 
  \delta^{(1,0)}_{{\cal Y}}Q - \delta^{(1,0)}_{{\cal X}}Q &=& 
  {\pounds}_{\xi_{(1,0)}}Q_{0}, \\
  \label{eq:Bruni-48} 
  \delta^{(0,1)}_{\cal Y}Q - \delta^{(0,1)}_{\cal X}Q &=& 
  {\pounds}_{\xi_{(0,1)}}Q_{0}, \\
  \label{eq:Bruni-49} 
  \delta^{(2,0)}_{\cal Y}Q - \delta^{(2,0)}_{\cal X}Q &=& 
  2 {\pounds}_{\xi_{(1,0)}} \delta^{(1,0)}_{\cal X}Q 
  +\left\{{\pounds}_{\xi_{(2,0)}}+{\pounds}_{\xi_{(1,0)}}^{2}\right\} Q_{0},\\
  \label{eq:Bruni-50} 
  \delta^{(1,1)}_{\cal Y}Q - \delta^{(1,1)}_{\cal X}Q &=& 
  {\pounds}_{\xi_{(1,0)}} \delta^{(0,1)}_{\cal X}Q 
  + {\pounds}_{\xi_{(0,1)}} \delta^{(1,0)}_{\cal X}Q 
  \nonumber\\
  && 
  + \left\{{\pounds}_{\xi_{(1,1)}} 
    + \frac{1}{2} {\pounds}_{\xi_{(1,0)}}{\pounds}_{\xi_{(0,1)}}
    + \frac{1}{2} {\pounds}_{\xi_{(0,1)}}{\pounds}_{\xi_{(1,0)}}
  \right\} Q_{0},\\
  \label{eq:Bruni-51} 
  \delta^{(0,2)}_{\cal Y}Q - \delta^{(0,2)}_{\cal X}Q &=& 
  2 {\pounds}_{\xi_{(0,1)}} \delta^{(0,1)}_{\cal X}Q 
  +\left\{{\pounds}_{\xi_{(0,2)}}+{\pounds}_{\xi_{(0,1)}}^{2}\right\} Q_{0},\\
  \label{eq:Bruni-52} 
  \delta^{(3,0)}_{\cal Y}Q - \delta^{(3,0)}_{\cal X}Q &=& 
  3 {\pounds}_{\xi_{(1,0)}}\delta^{(2,0)}_{\cal X}Q 
  + 3 \left\{ {\pounds}_{\xi_{(2,0)}} + {\pounds}_{\xi_{(1,0)}}^{2} \right\}
  \delta^{(1,0)}_{\cal X}Q 
  \nonumber\\
  && 
  + \left\{ {\pounds}_{\xi_{(3,0)}} 
    + 3 {\pounds}_{\xi_{(1,0)}}{\pounds}_{\xi_{(2,0)}} 
    + {\pounds}_{\xi_{(1,0)}}^{3} \right\} Q_{0}, \\ 
  \label{eq:Bruni-53} 
  \delta^{(2,1)}_{\cal Y}Q - \delta^{(2,1)}_{\cal X}Q &=& 
  2 {\pounds}_{\xi_{(1,0)}}\delta^{(1,1)}_{\cal X}Q 
  + {\pounds}_{\xi_{(0,1)}}\delta^{(2,0)}_{\cal X}Q 
  \nonumber\\
  && 
  + \left\{ {\pounds}_{\xi_{(2,0)}} + {\pounds}_{\xi_{(1,0)}}^{2} \right\} 
  \delta^{(0,1)}_{\cal X}Q 
  \nonumber\\
  && 
  + 2 \left\{ {\pounds}_{\xi_{(1,1)}} 
    + \frac{1}{2} {\pounds}_{\xi_{(1,0)}} {\pounds}_{\xi_{(0,1)}} 
    + \frac{1}{2} {\pounds}_{\xi_{(0,1)}} {\pounds}_{\xi_{(1,0)}} 
  \right\} \delta^{(1,0)}_{\cal X}Q 
  \nonumber\\
  && 
  + \left\{ {\pounds}_{\xi_{(2,1)}} 
    + 2 {\pounds}_{\xi_{(1,0)}}{\pounds}_{\xi_{(1,1)}} 
    + {\pounds}_{\xi_{(0,1)}} {\pounds}_{\xi_{(2,0)}} 
  \right.
  \nonumber\\
  && \quad\quad\quad 
  \left.
    + {\pounds}_{\xi_{(1,0)}} {\pounds}_{\xi_{(0,1)}}
    {\pounds}_{\xi_{(1,0)}}  
  \right\} Q_{0}, \\ 
  \label{eq:Bruni-54} 
  \delta^{(1,2)}_{\cal Y}Q - \delta^{(1,2)}_{\cal X}Q &=& 
  2 {\pounds}_{\xi_{(0,1)}}\delta^{(1,1)}_{\cal X}Q 
  + {\pounds}_{\xi_{(1,0)}}\delta^{(0,2)}_{\cal X}Q 
  \nonumber\\
  && 
  + \left\{ {\pounds}_{\xi_{(0,2)}} + {\pounds}_{\xi_{(0,1)}}^{2} \right\} 
  \delta^{(1,0)}_{\cal X}Q 
  \nonumber\\
  && 
  + 2 \left\{ {\pounds}_{\xi_{(1,1)}} 
    + \frac{1}{2} {\pounds}_{\xi_{(1,0)}} {\pounds}_{\xi_{(0,1)}} 
    + \frac{1}{2} {\pounds}_{\xi_{(0,1)}} {\pounds}_{\xi_{(1,0)}} 
  \right\} \delta^{(0,1)}_{\cal X}Q 
  \nonumber\\
  && 
  + \left\{ {\pounds}_{\xi_{(1,2)}} 
    + 2 {\pounds}_{\xi_{(0,1)}}{\pounds}_{\xi_{(1,1)}} 
    + {\pounds}_{\xi_{(1,0)}} {\pounds}_{\xi_{(0,2)}} 
  \right.
  \nonumber\\
  && \quad\quad\quad 
  \left.
    + {\pounds}_{\xi_{(0,1)}} {\pounds}_{\xi_{(1,0)}} {\pounds}_{\xi_{(0,1)}}  
  \right\} Q_{0}, \\ 
  \label{eq:Bruni-55} 
  \delta^{(0,3)}_{\cal Y}Q - \delta^{(0,3)}_{\cal X}Q &=& 
  3 {\pounds}_{\xi_{(0,1)}}\delta^{(0,2)}_{\cal X}Q 
  + 3 \left\{ {\pounds}_{\xi_{(0,2)}} + {\pounds}_{\xi_{(0,1)}}^{2} \right\}
  \delta^{(0,1)}_{\cal X}Q 
  \nonumber\\
  && 
  + \left\{ {\pounds}_{\xi_{(0,3)}} 
    + 3 {\pounds}_{\xi_{(0,1)}}{\pounds}_{\xi_{(0,2)}} 
    + {\pounds}_{\xi_{(0,1)}}^{3} \right\} Q_{0}.
\end{eqnarray}
These results are, of course, consistent with the gauge
invariance up to order ($n,n'$) as introduced above.
Equation (\ref{eq:Bruni-47}) [or (\ref{eq:Bruni-48})] implies
that $Q_{\lambda,\epsilon}$ is gauge invariant up to order
$(1,0)$ [or $(0,1)$] iff ${\pounds}_{\xi}Q_{0}=0$ for any
vector field $\xi^{a}$ on ${\cal M}_{0}$.
Equation (\ref{eq:Bruni-49}) implies that
$Q_{\lambda,\epsilon}$ is gauge invariant up to order $(2,0)$
iff ${\pounds}_{\xi}Q_{0}=0$ {\it and}
${\pounds}_{\xi}\delta^{(1,0)}_{\cal X}Q_{0}=0$ for any
vector field on ${\cal M}_{0}$.
This can be repreated analogously for all the orders.

%************************************************

%%%%%%%%%%%%%%%%%%%%%%%%%%%%%%%%%%%%%%%%%%%%%%%%%%%%%%%%%%%%%%%%%%%%%
\subsection{Coordinate transformations}
%%%%%%%%%%%%%%%%%%%%%%%%%%%%%%%%%%%%%%%%%%%%%%%%%%%%%%%%%%%%%%%%%%%%%

%************************************************

The above formulation of the perturbations and their gauge
transformation are independent of the explicit form of the
coordinate system. 
In some situations, it is convenient to introduce an explicit
coordinate system in order to carry out explicit calculations in
a practical case. 
Then, it is instructive to show the above geometrical
formulation of gauge transformations in terms of the
corresponding coordinate transformations, though it is not
necessary for the discussion in \S\S4 and 5. 
Here, we briefly discuss this coordinate transformation.
The explicit forms of this transformation given below show the
fact that the gauge freedom in perturbation theory is more than
the usual assignment of coordinate labels to points on the
single spacetime.
Details of the outline given here are given in the series of
papers by Bruni and coworkers
\cite{Bruni-Gualtieri-Sopuerta,Matarrese-Mollerach-Bruni}.

%************************************************

We have considered two gauge choices, represented  by the
groups of diffeomorphisms ${\cal X}_{\lambda,\epsilon}$ and
${\cal Y}_{\lambda,\epsilon}$, under which a point $o$ on the
physical spacetime ${\cal M}_{\lambda,\epsilon}$ corresponds to
two different points in the background manifold ${\cal M}_{0}$:
$p={\cal X}^{-1}_{\lambda,\epsilon}(o)$ and 
$q={\cal Y}^{-1}_{\lambda,\epsilon}(o)$ as depicted in Fig.
\ref{fig:figure1}.
The transformation relating these two gauge choices is
described by the two-parameter family of diffeomorphisms 
$\Phi_{\lambda,\epsilon} = {\cal X}^{-1}_{\lambda,\epsilon}
\circ {\cal Y}_{\lambda,\epsilon}$, 
so that $\Phi_{\lambda,\epsilon}(q)=p$.
Under this gauge transformation, a tensor field $Q$ on
$p\in{\cal M}_{0}$ is pulled back to the tensor field
$(\Phi^{*}Q)(q) = \Phi^{*}(Q(p))$ on $q\in{\cal M}_{0}$.

%************************************************

Now, let us consider a chart $({\cal U},X)$ on an open
subset ${\cal U}$ of the background ${\cal M}_{0}$.
The coordinate system $X$ is a map from the manifold 
${\cal M}_{0}$ to $\MR^{m}$.
Since the gauges ${\cal X}_{\lambda,\epsilon}$ and
${\cal Y}_{\lambda,\epsilon}$ are maps from the background
${\cal M}_{0}$ to the physical spacetime 
${\cal M}_{\lambda,\epsilon}$,
these gauges define two maps from ${\cal M}_{\lambda,\epsilon}$
to $\MR^{m}$: 
\begin{eqnarray}
  X\circ{\cal X}^{-1}_{\lambda,\epsilon} : 
  {\cal M}_{\lambda,\epsilon} &\rightarrow& \MR^{m} \quad\quad\quad\quad
  X\circ{\cal Y}^{-1}_{\lambda,\epsilon} : 
  {\cal M}_{\lambda,\epsilon} \rightarrow \MR^{m} \nonumber\\
  o &\mapsto& x(p(o)), 
  \quad\quad\quad\quad\quad\quad\quad\quad\;\;
  o \mapsto x(q(o)). 
\end{eqnarray}
The gauge transformation $\Phi_{\lambda,\epsilon}$ is
regarded as the transformation of these maps.

%************************************************

It is well-known that there are two different points of view with
which we can regard the gauge transformation
$\Phi_{\lambda,\epsilon}$ as a change of the coordinate system,
the {\it active} point of view and the {\it passive} point of
view. 
In the {\it active} point of view, one considers a
diffeomorphism that changes the point on the background 
${\cal M}_{0}$, and the coordinate change 
$x^{\mu}(p) \rightarrow \tilde{x}^{\mu}(p)$ is given according
to the definition of the pull-back of $x$ as 
\begin{equation}
  \tilde{x}^{\mu}(p):=x^{\mu}(\Phi(p)) 
\end{equation}
By contrast, in the {\it passive} point of view, we introduce a
new chart $({\cal U}',Y)$ defined by
$Y:=X\circ\Phi^{-1}_{\lambda,\epsilon}$, and the two sets of
coordinates are related by
\begin{equation}
  y^{\mu}(p) = x^{\mu}(q).
\end{equation}
In this {\it passive} point of view, the gauge transformation is
regarded as not changing the point on ${\cal M}_{0}$ but changing
the chart from $({\cal U},X)$ to $({\cal U}',Y)$ 
(i.e. changing the labels of the points on ${\cal M}_{0}$).
The coordinate transformation is given by 
$x^{\mu}(q) \rightarrow y^{\mu}(q)$.

%************************************************

Now, let us consider the transformation of a vector field
$V^{a}$ and the coordinate transformation from the {\it active} and
{\it passive} points of view.

%************************************************

From the {\it active} point of view, the components $V^{\mu}$ of
a vector field $V^{a}$ in the chart $({\cal U},X)$ are related to
the components $\tilde{V}^{\mu}$ of the transformed vector
field $\tilde{V}^{a}$ as
\begin{equation}
  \tilde{V}^{\mu} = (X_{*}\tilde{V})^{*} =
  (X_{*}\Phi^{*}_{\lambda,\epsilon}V)^{\mu}. 
\end{equation}
In order to write down explicit expressions, we apply the
expansion of the pull-back of $\Phi^{*}_{\lambda,\epsilon}$ [see 
Eq.~(\ref{eq:two-parameter-Bruni-30-simpler})] to the coordinate 
functions $x^{\mu}$. 
Then, the {\it active} coordinate transformation is given by 
\begin{eqnarray}
  \tilde{x}^{\mu}(p) &=& x^{\mu}(q) = (\Phi^{*}x^{\mu})(p) \nonumber\\
  &=& x^{\mu}(p) + \lambda \xi^{\mu}_{(1,0)} + \epsilon \xi_{(0,1)} \nonumber\\
  && \quad + \frac{\lambda^{2}}{2}\left(\xi^{\mu}_{(2,0)} +
    \xi^{\nu}_{(1,0)}\partial_{\nu}\xi^{\mu}_{(1,0)} \right)
  + \frac{\epsilon^{2}}{2}\left(\xi^{\mu}_{(0,2)} +
    \xi^{\nu}_{(0,1)}\partial_{\nu}\xi^{\mu}_{(0,1)} \right) \nonumber\\
  && \quad + \lambda\epsilon \left(\xi^{\mu}_{(1,1)} 
    + \frac{1}{2}\xi^{\nu}_{(1,0)} \partial_{\nu}\xi^{\mu}_{(0,1)} 
    + \frac{1}{2}\xi^{\nu}_{(0,1)} \partial_{\nu}\xi^{\mu}_{(1,0)} 
  \right) \nonumber\\
  && \quad + \frac{\lambda^{3}}{6} \left(\xi^{\mu}_{(3,0)} 
    + 3 \xi^{\nu}_{(1,0)}\partial_{\nu}\xi^{\mu}_{(2,0)} 
    + \xi_{(1,0)}^{\rho}\partial_{\rho}\xi_{(1,0)}^{\nu}
    \partial_{\nu}\xi_{(1,0)}^{\mu}  
  \right) 
  \nonumber\\
  && \quad 
  + \frac{\lambda^{2}\epsilon}{2} \left(
    \xi^{\mu}_{(2,1)} 
    + 2 \xi^{\nu}_{(1,0)}\partial_{\nu}\xi^{\mu}_{(1,1)}
    + \xi^{\nu}_{(0,1)}\partial_{\nu}\xi^{\mu}_{(2,0)}
  \right.
  \nonumber\\
  && \quad\quad\quad\quad 
  \left.
    + \xi^{\rho}_{(1,0)}
    \partial_{\rho}\xi^{\nu}_{(0,1)}\partial_{\nu}\xi^{\mu}_{(1,0)} 
  \right) 
  \nonumber\\
  && \quad 
  + \frac{\lambda\epsilon^{2}}{2} \left(
    \xi^{\mu}_{(1,2)} 
    + 2 \xi^{\nu}_{(0,1)}\partial_{\nu}\xi^{\mu}_{(1,1)}
    + \xi^{\nu}_{(1,0)}\partial_{\nu}\xi^{\mu}_{(0,2)}
  \right.
  \nonumber\\
  && \quad\quad\quad\quad 
  \left.
    + \xi^{\rho}_{(0,1)}
    \partial_{\rho}\xi^{\nu}_{(1,0)}\partial_{\nu}\xi^{\mu}_{(0,1)} 
  \right)
  \nonumber\\
  && \quad 
  + \frac{\epsilon^{3}}{6} \left(\xi^{\mu}_{(0,3)} 
    + 3 \xi^{\nu}_{(0,1)}\partial_{\nu}\xi^{\mu}_{(0,2)} 
    + \xi_{(0,1)}^{\rho}\partial_{\rho}\xi_{(0,1)}^{\nu}
    \partial_{\nu}\xi_{(0,1)}^{\mu}  
  \right) 
  \nonumber\\
  && \quad 
  +O^{4}(\lambda,\epsilon),
  \label{eq:Bruni-75}
\end{eqnarray}
where the vector fields $\xi^{\mu}_{(p,q)}$ and their
derivatives are evaluated in $x(p)$.
This expression gives the relation between the
coordinates, in the chart $({\cal U},X)$, of the two different points
$p$ and $q$ of ${\cal M}_{0}$.

%************************************************

From the {\it passive} point of view, we can use the properties
relating the pull-back and push-forward maps associated with
diffeomorphisms,  
\begin{equation}
  X_{*}\Phi^{*}_{\lambda,\epsilon}V =
  X_{*}\Phi_{*\lambda,\epsilon}^{-1}V = Y_{*}V.
\end{equation}
Thus, we obtain the well-known result that the components of the
transformed vector $\tilde{V}^{a}$ in the coordinate system $X$
are defined in terms of the components of the vector $V^{a}$ in
the new coordinate system $Y$:
\begin{equation}
  \tilde{V}^{\mu}(x(p)) = \left(Y_{*}V(q)\right)^{\mu} =
  V'^{\mu}(y(q)) = \left.\left(
      \frac{\partial y^{\mu}}{\partial x^{\nu}}
    \right)\right|_{x(q)}V^{\nu}(x(q)).
  \label{eq:Bruni-74}
\end{equation}

%************************************************

In order to write down explicit expressions, we apply the
expansion of the pull-back of $\Phi^{*}_{\lambda,\epsilon}$ [see 
Eq.~(\ref{eq:two-parameter-Bruni-30-simpler})] to the coordinate 
functions $x^{\mu}$.
The {\it passive} coordinate transformation is found by
inverting (\ref{eq:Bruni-75}).
Using the representation
(\ref{eq:upto-third-ordered-gauge-transformation-X-Y-simpler})
of the Taylor expansion of the inverse of
$\Phi_{\lambda,\epsilon}^{*}$, we obtain the {\it passive}
coordinate transformation:
\begin{eqnarray}
  y^{\mu}(q) &:=& x^{\mu}(p) 
  = \left(\left(\Phi^{-1}\right)^{*}x^{\mu}\right)(q) \nonumber\\
  &=& x^{\mu}(q)
  - \lambda \xi_{(1,0)}^{\mu}(q)
  - \epsilon \xi_{(0,1)}^{\mu}(q)
  \nonumber\\
  &&
  + \frac{\lambda^{2}}{2} \left\{
    - \xi_{(2,0)}^{\mu}(q)
    + \xi_{(1,0)}^{\nu}(q) \partial_{\nu} \xi_{(1,0)}^{\mu}(q)
  \right\}
  \nonumber\\
  && 
  + \frac{\epsilon^{2}}{2} \left\{
    - \xi_{(0,2)}^{\mu}(q) 
    + \xi_{(0,1)}^{\nu}(q) \partial_{\nu} \xi_{(0,1)}^{\mu}(q)
  \right\}
  \nonumber\\
  && 
  + \lambda\epsilon \left\{ 
    - \xi_{(1,1)}^{\mu}(q)
    + \frac{1}{2} \xi_{(1,0)}^{\nu}(q) \partial_{\nu} \xi_{(0,1)}^{\mu}(q)
    + \frac{1}{2} \xi_{(0,1)}^{\nu}(q) \partial_{\nu} \xi_{(1,0)}^{\mu}(q)
  \right\}
  \nonumber\\
  &&
  + \frac{\lambda^{3}}{6} \left\{
    - \xi_{(3,0)}^{\mu}(q)
    + 3 \xi_{(2,0)}^{\nu}(q) \partial_{\nu} \xi_{(1,0)}^{\mu}(q)
  \right.
  \nonumber\\
  && \quad\quad\quad
  \left.
    - \xi_{(1,0)}^{\rho}(q) \partial_{\rho} \left(
      \xi_{(1,0)}^{\nu}(q) \partial_{\nu} \xi_{(1,0)}^{\mu}(q)
    \right)
  \right\}
  \nonumber\\
  &&
  + \frac{\lambda^{2}\epsilon}{2} \left\{
    - \xi_{(2,1)}^{\mu}(q)
    + \xi_{(2,0)}^{\nu}(q) \partial_{\nu} \xi_{(0,1)}^{\mu}(q)
    + 2 \xi_{(1,1)}^{\nu}(q) \partial_{\nu} \xi_{(1,0)}^{\mu}(q)
  \right.
  \nonumber\\
  &&\quad\quad\quad
  \left.
    - \xi_{(1,0)}^{\rho}(q) \partial_{\rho} \left(
      \xi_{(0,1)}^{\nu}(q) \partial_{\nu} \xi_{(1,0)}^{\mu}(q)
    \right)
  \right\}
  \nonumber\\
  &&
  + \frac{\lambda\epsilon^{2}}{2} \left\{
    - \xi_{(1,2)}^{\mu}(q)
    + \xi_{(0,2)}^{\nu}(q) \partial_{\nu} \xi_{(1,0)}^{\mu}(q)
    + 2 \xi_{(1,1)}^{\nu}(q) \partial_{\nu} \xi_{(0,1)}^{\mu}(q)
  \right.
  \nonumber\\
  &&\quad\quad\quad
  \left.
    - \xi_{(0,1)}^{\rho}(q) \partial_{\rho} \left(
      \xi_{(1,0)}^{\nu}(q) \partial_{\nu} \xi_{(0,1)}^{\mu}(q)
    \right)
  \right\}
  \nonumber\\
  &&
  + \frac{\epsilon^{3}}{6} \left\{
    - \xi_{(0,3)}^{\mu}(q) 
    + 3 \xi_{(0,2)}^{\nu}(q) \partial_{\nu} \xi_{(0,1)}^{\mu}(q)
  \right.
  \nonumber\\
  && \quad\quad\quad
  \left.
    - \xi_{(0,1)}^{\rho}(q) \partial_{\rho} \left(
      \xi_{(0,1)}^{\nu}(q) \partial_{\nu} \xi_{(0,1)}(q)
    \right)
  \right\}
  \nonumber\\
  &&
  + O^{4}(\lambda,\epsilon).
  \label{eq:passive-coordinate-transformation}
\end{eqnarray}
This gives the relation between the coordinates of any arbitrary
point $q\in{\cal M}_{0}$ in the two different charts $({\cal U},X)$
and $({\cal U}',Y)$.

%************************************************

%%%%%%%%%%%%%%%%%%%%%%%%%%%%%%%%%%%%%%%%%%%%%%%%%%%%%%%%%%%%%%%%%%%%%
\section{Gauge invariant variables of higher order perturbations}
\label{sec:gauge-invariant}
%%%%%%%%%%%%%%%%%%%%%%%%%%%%%%%%%%%%%%%%%%%%%%%%%%%%%%%%%%%%%%%%%%%%%

%************************************************

Now, we consider the definitions of gauge invariant variables
for the perturbations. 
The gauge invariance we consider here is that up to order
($n,n'$) for ($n,n'$)-order perturbations as mentioned in
\S\ref{sec:Bruni-gauge-inv-gauge-trans}.
We do this because the gauge invariance to all orders is not so
useful\cite{Bruni-Gualtieri-Sopuerta,M.Bruni-S.Soonego-CQG1997,M.Bruni-S.Sonego-CQG-1999}.
Of course, the definition of gauge invariant variables is not
unique, because any function of gauge invariant variables
is also gauge invariant.
In this section, we present one procedure to define gauge
invariant variables.
To do this, we first consider the procedure to obtain the gauge
invariant variables for metric perturbations.
Next, we extend the procedure to define gauge invariant
variables for any physical variables, other than the metric.

%************************************************

%%%%%%%%%%%%%%%%%%%%%%%%%%%%%%%%%%%%%%%%%%%%%%%%%%%%%%%%%%%%%%%%%%%%%
\subsection{Metric perturbations} 
%%%%%%%%%%%%%%%%%%%%%%%%%%%%%%%%%%%%%%%%%%%%%%%%%%%%%%%%%%%%%%%%%%%%%

%************************************************

As seen in Eqs.~(\ref{eq:Bruni-39}) and (\ref{eq:Bruni-40}), we
first consider the Taylor expansion of the spacetime metric
$g_{ab}(\lambda,\epsilon)$ on ${\cal M}_{\lambda,\epsilon}$.
As discuss in \S\ref{sec:Bruni-gauge-inv-gauge-trans}, the
Taylor expansion of the spacetime metric up to third order is
carried out by using a gauge choice ${\cal X}$ on the background
spacetime ${\cal M}_{0}$:  
\begin{eqnarray}
  {}^{\cal X}g_{ab} &=& {}^{(0)}g_{ab}
  + \sum^{\infty}_{k,k'=1} \frac{\lambda^{k}\epsilon^{k'}}{k!k'!} 
  {}^{(k,k')}_{\quad\;{\cal X}}h_{ab}
  \nonumber\\
  &=& {}^{(0)}g_{ab} 
  + \lambda {}^{(1,0)}_{\quad{\cal X}}h_{ab}
  + \epsilon {}^{(0,1)}_{\quad{\cal X}}h_{ab}
  + \frac{\lambda^{2}}{2} {}^{(2,0)}_{\quad{\cal X}}h_{ab} 
  + \lambda\epsilon {}^{(1,1)}_{\quad{\cal X}}h_{ab}
  + \frac{\epsilon^{2}}{2} {}^{(0,2)}_{\quad{\cal X}}h_{ab}
  \nonumber\\
  && \quad
  + \frac{\lambda^{3}}{3!} {}^{(3,0)}_{\quad{\cal X}}h_{ab} 
  + \frac{\lambda^{2}\epsilon}{2} {}^{(2,1)}_{\quad{\cal X}}h_{ab}
  + \frac{\lambda\epsilon^{2}}{2} {}^{(1,2)}_{\quad{\cal X}}h_{ab}
  + \frac{\epsilon^{3}}{3!} {}^{(0,3)}_{\quad{\cal X}}h_{ab}
  \nonumber\\
  && \quad
  + O(\lambda,\epsilon)^{4}.
\end{eqnarray}

%************************************************

From Eqs.~(\ref{eq:Bruni-47})--(\ref{eq:Bruni-55}), the gauge
transformation rules for the metric perturbations up to third order
are given by 
\begin{eqnarray}
  \label{eq:metric-gauge-trans-1.0} 
  {}^{(1,0)}_{\quad{\cal Y}}h_{ab} - {}^{(1,0)}_{\quad{\cal X}}h_{ab}
  &=& {\pounds}_{\xi_{(1,0)}}{}^{(0)}g_{ab}, \\
  \label{eq:metric-gauge-trans-0.1}
  {}^{(0,1)}_{\quad{\cal Y}}h_{ab} - {}^{(0,1)}_{\quad{\cal X}}h_{ab}
  &=& 
  {\pounds}_{\xi_{(0,1)}}{}^{(0)}g_{ab}, \\
  \label{eq:metric-gauge-trans-2.0}
  {}^{(2,0)}_{\quad{\cal Y}}h_{ab} - {}^{(2,0)}_{\quad{\cal X}}h_{ab}
  &=& 
  2 {\pounds}_{\xi_{(1,0)}} {}^{(1,0)}_{\quad{\cal X}}h_{ab} 
  + \left\{
    {\pounds}_{\xi_{(2,0)}} + {\pounds}_{\xi_{(1,0)}}^{2}
  \right\} {}^{(0)}g_{ab},\\
  \label{eq:metric-gauge-trans-1.1}
  {}^{(1,1)}_{\quad{\cal Y}}h_{ab} - {}^{(1,1)}_{\quad{\cal X}}h_{ab}
  &=& 
  {\pounds}_{\xi_{(1,0)}} {}^{(0,1)}_{\quad{\cal X}}h_{ab} 
  + {\pounds}_{\xi_{(0,1)}} {}^{(1,0)}_{\quad{\cal X}}h_{ab} 
  \nonumber\\
  && 
  + \left\{{\pounds}_{\xi_{(1,1)}} 
    + \frac{1}{2} {\pounds}_{\xi_{(1,0)}}{\pounds}_{\xi_{(0,1)}}
    + \frac{1}{2} {\pounds}_{\xi_{(0,1)}}{\pounds}_{\xi_{(1,0)}}
  \right\} {}^{(0)}g_{ab},\\
  \label{eq:metric-gauge-trans-0.2}
  {}^{(0,2)}_{\quad{\cal Y}}h_{ab} - {}^{(0,2)}_{\quad{\cal X}}h_{ab}
  &=& 
  2 {\pounds}_{\xi_{(0,1)}} {}^{(0,1)}_{\quad{\cal X}}h_{ab} 
  +\left\{{\pounds}_{\xi_{(0,2)}}+{\pounds}_{\xi_{(0,1)}}^{2}\right\} {}^{(0)}g_{ab},\\
  \label{eq:metric-gauge-trans-3.0}
  {}^{(3,0)}_{\quad{\cal Y}}h_{ab} - {}^{(3,0)}_{\quad{\cal X}}h_{ab}
  &=& 
  3 {\pounds}_{\xi_{(1,0)}}{}^{(2,0)}_{\quad{\cal X}}h_{ab}
  + 3 \left\{ {\pounds}_{\xi_{(2,0)}} + {\pounds}_{\xi_{(1,0)}}^{2} \right\}
  {}^{(1,0)}_{\quad{\cal X}}h_{ab} 
  \nonumber\\
  && 
  + \left\{ {\pounds}_{\xi_{(3,0)}} 
    + 3 {\pounds}_{\xi_{(1,0)}}{\pounds}_{\xi_{(2,0)}} 
    + {\pounds}_{\xi_{(1,0)}}^{3} \right\} {}^{(0)}g_{ab}, \\ 
  \label{eq:metric-gauge-trans-2.1}
  {}^{(2,1)}_{\quad{\cal Y}}h_{ab} - {}^{(2,1)}_{\quad{\cal X}}h_{ab}
  &=& 
  2 {\pounds}_{\xi_{(1,0)}}{}^{(1,1)}_{\quad{\cal X}}h_{ab}
  + {\pounds}_{\xi_{(0,1)}}{}^{(2,0)}_{\quad{\cal X}}h_{ab} 
  \nonumber\\
  && 
  + \left\{ {\pounds}_{\xi_{(2,0)}} + {\pounds}_{\xi_{(1,0)}}^{2} \right\} 
  {}^{(0,1)}_{\quad{\cal X}}h_{ab} 
  \nonumber\\
  && 
  + 2 \left\{ {\pounds}_{\xi_{(1,1)}} 
    + \frac{1}{2} {\pounds}_{\xi_{(1,0)}} {\pounds}_{\xi_{(0,1)}} 
    + \frac{1}{2} {\pounds}_{\xi_{(0,1)}} {\pounds}_{\xi_{(1,0)}} 
  \right\} {}^{(1,0)}_{\quad{\cal X}}h_{ab} 
  \nonumber\\
  && 
  + \left\{ {\pounds}_{\xi_{(2,1)}} 
    + 2 {\pounds}_{\xi_{(1,0)}}{\pounds}_{\xi_{(1,1)}} 
    + {\pounds}_{\xi_{(0,1)}} {\pounds}_{\xi_{(2,0)}} 
  \right.
  \nonumber\\
  && \quad\quad\quad
  \left.
    + {\pounds}_{\xi_{(1,0)}} {\pounds}_{\xi_{(0,1)}}
    {\pounds}_{\xi_{(1,0)}}  
  \right\} {}^{(0)}g_{ab}, \\ 
  \label{eq:metric-gauge-trans-1.2}
  {}^{(1,2)}_{\quad{\cal Y}}h_{ab} - {}^{(1,2)}_{\quad{\cal X}}h_{ab}
  &=& 
  2 {\pounds}_{\xi_{(0,1)}}{}^{(1,1)}_{\quad{\cal X}}h_{ab} 
  + {\pounds}_{\xi_{(1,0)}}{}^{(0,2)}_{\quad{\cal X}}h_{ab} 
  \nonumber\\
  && 
  + \left\{ {\pounds}_{\xi_{(0,2)}} + {\pounds}_{\xi_{(0,1)}}^{2} \right\} 
  {}^{(1,0)}_{\quad{\cal X}}h_{ab} 
  \nonumber\\
  && 
  + 2 \left\{ {\pounds}_{\xi_{(1,1)}} 
    + \frac{1}{2} {\pounds}_{\xi_{(1,0)}} {\pounds}_{\xi_{(0,1)}} 
    + \frac{1}{2} {\pounds}_{\xi_{(0,1)}} {\pounds}_{\xi_{(1,0)}} 
  \right\} {}^{(0,1)}_{\quad{\cal X}}h_{ab} 
  \nonumber\\
  && 
  + \left\{ {\pounds}_{\xi_{(1,2)}} 
    + 2 {\pounds}_{\xi_{(0,1)}}{\pounds}_{\xi_{(1,1)}} 
    + {\pounds}_{\xi_{(1,0)}} {\pounds}_{\xi_{(0,2)}} 
  \right.
  \nonumber\\
  && \quad\quad\quad
  \left.
    + {\pounds}_{\xi_{(0,1)}} {\pounds}_{\xi_{(1,0)}} {\pounds}_{\xi_{(0,1)}}  
  \right\} {}^{(0)}g_{ab}, \\ 
  \label{eq:metric-gauge-trans-0.3}
  {}^{(0,3)}_{\quad{\cal Y}}h_{ab} - {}^{(0,3)}_{\quad{\cal X}}h_{ab}
  &=& 
  3 {\pounds}_{\xi_{(0,1)}}{}^{(0,2)}_{\quad{\cal X}}h_{ab} 
  + 3 \left\{ {\pounds}_{\xi_{(0,2)}} + {\pounds}_{\xi_{(0,1)}}^{2} \right\}
  {}^{(0,1)}_{\quad{\cal X}}h_{ab} 
  \nonumber\\
  && 
  + \left\{ {\pounds}_{\xi_{(0,3)}} 
    + 3 {\pounds}_{\xi_{(0,1)}}{\pounds}_{\xi_{(0,2)}} 
    + {\pounds}_{\xi_{(0,1)}}^{3} \right\} {}^{(0)}g_{ab}.
\end{eqnarray}
Inspecting these transformation rules, we consider the procedure
to separate the gauge invariant parts and gauge variant parts of
the metric perturbation at each order.
The aim of this paper is to show that this separation for higher
order perturbations can be carried out with the same procedure
as for linear perturbation theory. 
In other words, if we are able to carry out this separation at
linear order, the separation for higher order perturbations can
also be carried out.

%************************************************

%%%%%%%%%%%%%%%%%%%%%%%%%%%%%%%%%%%%%%%%%%%%%%%%%%%%%%%%%%%%%%%%%%%%%
\subsubsection{Linear order perturbations}

%************************************************

Suppose that, inspecting the gauge transformation rules
(\ref{eq:metric-gauge-trans-1.0}) and
(\ref{eq:metric-gauge-trans-0.1}), the $O(\lambda)$ and
$O(\epsilon)$ perturbations can be decomposed as   
\begin{eqnarray}
  {}^{(1,0)}_{\quad{\cal X}}h_{ab} 
  &=:& {}^{(1,0)}_{\quad{\cal X}}{\cal H}_{ab} 
  + \nabla_{a}{}^{(1,0)}_{\quad{\cal X}}X_{b} 
  + \nabla_{b}{}^{(1,0)}_{\quad{\cal X}}X_{a}, 
  \nonumber\\
  {}^{(0,1)}_{\quad{\cal X}}h_{ab} 
  &=:& {}^{(0,1)}_{\quad{\cal X}}{\cal H}_{ab} 
  + \nabla_{a}{}^{(0,1)}_{\quad{\cal X}}X_{b}
  + \nabla_{b}{}^{(0,1)}_{\quad{\cal X}}X_{a},
  \label{eq:linear-metric-decomp}
\end{eqnarray}
so that the variables ${}^{(1,0)}_{\quad{\cal X}}{\cal H}_{ab}$
and ${}^{(0,1)}_{\quad{\cal X}}{\cal H}_{ab}$ are gauge
invariant; i.e.,  
\begin{equation}
  {}^{(1,0)}_{\quad{\cal Y}}{\cal H}_{ab} 
  - {}^{(1,0)}_{\quad{\cal X}}{\cal H}_{ab} = 0,
  \quad 
  {}^{(0,1)}_{\quad{\cal Y}}{\cal H}_{ab} 
  - {}^{(0,1)}_{\quad{\cal X}}{\cal H}_{ab} = 0,
\end{equation}
under the gauge transformation $\Phi={\cal X}^{-1}\circ{\cal Y}$.
Next, note that the vector fields 
${}^{(1,0)}_{\quad{\cal X}}X_{a}$ and  
${}^{(0,1)}_{\quad{\cal X}}X_{a}$ are transformed as 
\begin{equation}
  \label{eq:(1,0)-and-(0,1)-X-a-gauge-trans}
  {}^{(1,0)}_{\quad{\cal Y}}X^{a} 
  - {}^{(1,0)}_{\quad{\cal X}}X^{a} =
  \xi^{a}_{(1,0)}, 
  \quad
  {}^{(0,1)}_{\quad{\cal Y}}X^{a} 
  - {}^{(0,1)}_{\quad{\cal X}}X^{a} =
  \xi^{a}_{(0,1)}; 
\end{equation}
i.e., ${}^{(1,0)}_{\quad{\cal X}}X^{a}$ and
${}^{(0,1)}_{\quad{\cal X}}X^{a}$ are the gauge variant parts of
the metric perturbations ${}^{(1,0)}_{\quad{\cal X}}h_{ab}$
and ${}^{(0,1)}_{\quad{\cal X}}h_{ab}$, respectively.
We also note that the number of independent components of 
${}^{(1,0)}_{\quad{\cal X}}{\cal H}_{ab}$ (or 
${}^{(0,1)}_{\quad{\cal X}}{\cal H}_{ab}$) is $m(m-1)/2$, where $m$ 
is the dimension of the spacetime manifold.

%************************************************

It is non-trivial to carry out the systematic decomposition
(\ref{eq:linear-metric-decomp}) on an arbitrary background
spacetime, and the procedure completely depends on the background
spacetime  $({\cal M}_{0},{}^{(0)}g_{ab})$. 
For simple background spacetimes in which there are some Killing
symmetries, one useful type of analyses to carry out the
decomposition (\ref{eq:linear-metric-decomp}) is that based on
the expansion in harmonic functions on a submanifold of
the background spacetime $({\cal M}_{0},{}^{(0)}g_{ab})$
\cite{Kodama-Sasaki,Gerlach_Sengupta,Chadrasekhar-book}.
For example, the harmonic functions on a homogeneous and isotropic
three-dimensional space are used in cosmological perturbation
theory\cite{Kodama-Sasaki}.
Analyses based on a harmonic expansion depend strongly on not
only local symmetries of the background spacetime but also
the global topology of the submanifold on which scalar, vector,
and tensor harmonics are defined.

%************************************************

In spite of the non-trivial nature of this decomposition, we
assume that the decomposition (\ref{eq:linear-metric-decomp})
can always be carried out in some manner. 
What is necessary for our discussion here is only the result of
the extraction of $m$ gauge variant components from $m(m+1)/2$
components of the metric perturbations 
${}^{(1,0)}_{\quad{\cal X}}h_{ab}$ or
${}^{(0,1)}_{\quad{\cal X}}h_{ab}$.
The details to realize the decomposition
(\ref{eq:linear-metric-decomp}) are not important to our 
discussion, though the existence of the procedure is crucial. 
As seen below, if such a procedure exists, we can easily show
that the procedure to define the gauge invariant variables of
higher order metric perturbations is reduced to that for linear
perturbations, whose existence we assume.
Though a generic formula to define gauge invariant variables
for any order perturbation might exist, we give only the
formulae for two-parameter perturbations up to third order.

%************************************************

%%%%%%%%%%%%%%%%%%%%%%%%%%%%%%%%%%%%%%%%%%%%%%%%%%%%%%%%%%%%%%%%%%%%%
\subsubsection{Second order perturbations}

%************************************************

Here, we consider the definitions of gauge invariant variables
for $O(\lambda^{2})$, $O(\epsilon^{2})$ and $O(\lambda\epsilon)$
metric perturbations.

%************************************************

First, we consider the $O(\lambda^{2})$ metric perturbation.
The metric perturbation ${}^{(2,0)}h_{ab}$ of this order is
transformed as in Eq.~(\ref{eq:metric-gauge-trans-2.0}) under
the gauge transformation $\Phi = {\cal X}^{-1}\circ{\cal Y}$.
Inspecting this transformation rule, we define a variable
${}^{(2,0)}_{\quad{\cal X}}\widehat{\cal H}_{ab}$ with the gauge 
${\cal X}$ by 
\begin{equation}
  \label{eq:widehat-H-ab-in-gauge-X-def-2.0}
  {}^{(2,0)}_{\quad{\cal X}}\widehat{\cal H}_{ab}
  := {}^{(2,0)}_{\quad{\cal X}}h_{ab}
  - 2 {\pounds}_{{}^{(1,0)}_{\quad{\cal X}}X} 
  {}^{(1,0)}_{\quad{\cal X}}h_{ab}
  + {\pounds}_{{}^{(1,0)}_{\quad{\cal X}}X}^{2} {}^{(0)}g_{ab}.
\end{equation}
We also define a variable 
${}^{(2,0)}_{\quad{\cal Y}}\widehat{\cal H}_{ab}$ with the gauge 
${\cal Y}$ by simply replacing ${\cal X}$ with ${\cal Y}$ in 
Eq.~(\ref{eq:widehat-H-ab-in-gauge-X-def-2.0}).
Using the gauge transformation rules
(\ref{eq:metric-gauge-trans-1.0}),
(\ref{eq:metric-gauge-trans-2.0}), and 
(\ref{eq:(1,0)-and-(0,1)-X-a-gauge-trans}), we can easily check
that the variable ${}^{(2,0)}_{\quad{\cal Y}}\widehat{\cal H}_{ab}$ is
transformed as 
\begin{eqnarray}
  \label{eq:widehat-metric-gauge-trans-2.0}
  {}^{(2,0)}_{\quad{\cal Y}}\widehat{\cal H}_{ab}
  -
  {}^{(2,0)}_{\quad{\cal X}}\widehat{\cal H}_{ab}
  = {\pounds}_{\sigma_{(2,0)}} {}^{(0)}g_{ab}, 
\end{eqnarray}
where the vector field $\sigma_{(2,0)}^{a}$ is defined by 
\begin{equation}
  \sigma_{(2,0)}^{a} := \xi_{(2,0)}^{a} 
  + [\xi_{(1,0)},{}^{(1,0)}_{\quad{\cal X}}X]^{a}.
\end{equation}
We note that the gauge transformation rule
(\ref{eq:widehat-metric-gauge-trans-2.0}) has the same form as the
gauge transformation rules (\ref{eq:metric-gauge-trans-1.0}) and
(\ref{eq:metric-gauge-trans-0.1}) of the linear metric perturbations.
Because we assume that the procedure to decompose a tensor field of
second rank, which transforms as
(\ref{eq:metric-gauge-trans-1.0}) or
(\ref{eq:metric-gauge-trans-0.1}), 
into the form (\ref{eq:linear-metric-decomp}) exists, the
transformation rule (\ref{eq:widehat-metric-gauge-trans-2.0})
implies that we can 
decompose the tensor  
${}^{(2,0)}_{\quad{\cal X}}\widehat{\cal H}_{ab}$ into a tensor
field ${}^{(2,0)}_{\quad{\cal X}}{\cal H}_{ab}$ and a vector field
${}^{(2,0)}_{\quad{\cal X}}X^{b}$ as
\begin{equation}
  \label{eq:(2,0)-order-metric-decomp}
  {}^{(2,0)}_{\quad{\cal X}}\widehat{\cal H}_{ab} =:
  {}^{(2,0)}_{\quad{\cal X}}{\cal H}_{ab} 
  + \nabla_{a}{}^{(2,0)}_{\quad{\cal X}}X_{b}
  + \nabla_{b}{}^{(2,0)}_{\quad{\cal X}}X_{a},
\end{equation}
where the tensor ${}^{(2,0)}_{\quad{\cal X}}{\cal H}_{ab}$ is gauge
invariant, i.e.,   
\begin{equation}
  {}^{(2,0)}_{\quad{\cal Y}}{\cal H}_{ab} - 
  {}^{(2,0)}_{\quad{\cal X}}{\cal H}_{ab} = 0,
\end{equation}
and the vector ${}^{(2,0)}_{\quad{\cal X}}X_{a}$ is gauge variant,
i.e., 
\begin{equation}
  \label{eq:(2,0)-X-a-gauge-trans}
  {}^{(2,0)}_{\quad{\cal Y}}X^{a} 
  - {}^{(2,0)}_{\quad{\cal X}}X^{a}
  = \xi_{(2,0)}^{a} + [\xi_{(1,0)},{}^{(1,0)}_{\quad{\cal X}}X]^{a}.
\end{equation}
Thus, we can extract the gauge invariant part 
${}^{(2,0)}{\cal H}_{ab}$ of the $O(\lambda^{2})$ metric
perturbation using the procedure to define the gauge invariant
variables for linear order metric perturbations.

%************************************************

The extraction of the gauge invariant part from the $O(\epsilon^{2})$
metric perturbation is accomplished in a manner that is completely
parallel to the above procedure for $O(\lambda^{2})$ perturbations. 
First, we define the variable 
${}^{(0,2)}_{\quad{\cal X}}\widehat{\cal H}_{ab}$
by 
\begin{equation}
  \label{eq:widehat-H-ab-in-gauge-X-def-0.2}
  {}^{(0,2)}_{\quad{\cal X}}\widehat{\cal H}_{ab}
  := {}^{(0,2)}_{\quad{\cal X}}h_{ab}
  - 2 {\pounds}_{{}^{(0,1)}_{\quad{\cal X}}X} 
  {}^{(0,1)}_{\quad{\cal X}}h_{ab}
  + {\pounds}_{{}^{(0,1)}_{\quad{\cal X}}X}^{2} {}^{(0)}g_{ab},
\end{equation}
with the gauge ${\cal X}$.
From the gauge transformation rules
(\ref{eq:metric-gauge-trans-0.1}), (\ref{eq:metric-gauge-trans-0.2})
and (\ref{eq:(1,0)-and-(0,1)-X-a-gauge-trans}), the variable 
${}^{(0,2)}_{\quad{\cal X}}\widehat{\cal H}_{ab}$ is transformed as 
\begin{eqnarray}
  \label{eq:widehat-metric-gauge-trans-0.2}
  {}^{(0,2)}_{\quad{\cal Y}}\widehat{\cal H}_{ab}
  - {}^{(0,2)}_{\quad{\cal X}}\widehat{\cal H}_{ab}
  = {\pounds}_{\sigma_{(0,2)}} {}^{(0)}g_{ab},
\end{eqnarray}
under the gauge transformation ${\cal X}\rightarrow{\cal Y}$,
where 
\begin{equation}
  \sigma_{(0,2)}^{a} := \xi_{(0,2)}^{a} 
  + [\xi_{(0,1)},{}^{(0,1)}_{\quad{\cal X}}X]^{a}.
\end{equation}
Because the gauge transformation rule
(\ref{eq:widehat-metric-gauge-trans-0.2}) has the same form as
that for the linear-order metric perturbations,
(\ref{eq:metric-gauge-trans-1.0}) and 
(\ref{eq:metric-gauge-trans-0.1}),
we can decompose the variable 
${}^{(0,2)}_{\quad{\cal X}}\widehat{\cal H}_{ab}$ as
\begin{equation}
  \label{eq:(0,2)-order-metric-decomp}
  {}^{(0,2)}_{\quad{\cal X}}\widehat{\cal H}_{ab} =:
  {}^{(0,2)}_{\quad{\cal X}}{\cal H}_{ab} 
  + \nabla_{a}{}^{(0,2)}_{\quad{\cal X}}X_{b}
  + \nabla_{b}{}^{(0,2)}_{\quad{\cal X}}X_{a},
\end{equation}
where ${}^{(0,2)}_{\quad{\cal X}}{\cal H}_{ab}$ is gauge
invariant, and the vector field ${}^{(0,2)}_{\quad{\cal X}}X_{a}$ is
the gauge variant part of the $O(\epsilon^{2})$ metric perturbations.
The vector field ${}^{(0,2)}_{\quad{\cal X}}X_{a}$ is transformed as
\begin{equation}
  \label{eq:(0,2)-X-a-gauge-trans}
  {}^{(0,2)}_{\quad{\cal Y}}X^{a} 
  - {}^{(0,2)}_{\quad{\cal X}}X^{a}
  = \xi_{(0,2)}^{a} + [\xi_{(0,1)},{}^{(0,1)}_{\quad{\cal X}}X]^{a}
\end{equation}
under the gauge transformation ${\cal X}\rightarrow{\cal Y}$.
Thus, we can find the gauge invariant part 
${}^{(0,2)}{\cal H}_{ab}$ of the $O(\epsilon^{2})$ metric
perturbation.

%************************************************

In a similar way, we can also define the gauge invariant
variables of the $O(\epsilon\lambda)$ metric perturbation
${}^{(1,1)}_{\quad{\cal X}}h_{ab}$.
The $O(\epsilon\lambda)$ metric perturbation is transformed as
in Eq.~(\ref{eq:metric-gauge-trans-1.1}) under the gauge
transformation ${\cal X}\rightarrow{\cal Y}$. 
Inspecting this gauge transformation rule, we first define the
variable ${}^{(1,1)}_{\quad{\cal X}}\widehat{\cal H}_{ab}$ by 
\begin{eqnarray}
  \label{eq:widehat-H-ab-in-gauge-X-def-1.1}
  {}^{(1,1)}_{\quad{\cal X}}\widehat{\cal H}_{ab}
  &:=& {}^{(1,1)}_{\quad{\cal X}}h_{ab}
  - {\pounds}_{{}^{(0,1)}_{\quad{\cal X}}X} {}^{(1,0)}_{\quad{\cal X}}h_{ab}
  - {\pounds}_{{}^{(1,0)}_{\quad{\cal X}}X} {}^{(0,1)}_{\quad{\cal X}}h_{ab}
  \nonumber\\
  && \quad
  + \frac{1}{2} \left\{
      {\pounds}_{{}^{(1,0)}_{\quad{\cal X}}X}
      {\pounds}_{{}^{(0,1)}_{\quad{\cal X}}X}
    + {\pounds}_{{}^{(0,1)}_{\quad{\cal X}}X}
      {\pounds}_{{}^{(1,0)}_{\quad{\cal X}}X}
  \right\} {}^{(0)}g_{ab}
\end{eqnarray}
in the gauge ${\cal X}$.
The variable ${}^{(1,1)}_{\quad{\cal Y}}\widehat{\cal H}_{ab}$ is
defined in the gauge ${\cal Y}$ in the same form as
Eq.~(\ref{eq:widehat-H-ab-in-gauge-X-def-1.1}), with the
replacement ${\cal X}\rightarrow{\cal Y}$. 
Using the gauge transformation rules
(\ref{eq:metric-gauge-trans-1.0}), (\ref{eq:metric-gauge-trans-0.1}), 
(\ref{eq:metric-gauge-trans-1.1}) and
(\ref{eq:(1,0)-and-(0,1)-X-a-gauge-trans}), we can easily check that
the variable ${}^{(1,1)}_{\quad{\cal X}}\widehat{\cal H}_{ab}$ is
transformed as 
\begin{eqnarray}
  \label{eq:widehat-metric-gauge-trans-1.1}
  {}^{(1,1)}_{\quad{\cal Y}}\widehat{\cal H}_{ab}
  - {}^{(1,1)}_{\quad{\cal X}}\widehat{\cal H}_{ab}
  = {\pounds}_{\sigma_{(1,1)}} {}^{(0)}g_{ab}, 
\end{eqnarray}
where the vector field $\sigma_{(1,1)}$ is defined by 
\begin{equation}
  \sigma_{(1,1)}^{a} = \xi_{(1,1)}^{a} 
  + \frac{1}{2}[\xi_{(0,1)}, {}^{(1,0)}_{\quad{\cal X}}X]^{a} 
  + \frac{1}{2}[\xi_{(1,0)}, {}^{(0,1)}_{\quad{\cal X}}X]^{a}.
\end{equation}
Here again, we have the gauge transformation rule
(\ref{eq:widehat-metric-gauge-trans-1.1}), which has the same
form as that for linear perturbations given in
(\ref{eq:metric-gauge-trans-1.0}) and
(\ref{eq:metric-gauge-trans-0.1}).
Then, as in the previous cases of $(2,0)$ and $(0,2)$ order
perturbations, we can decompose the variables 
${}^{(1,1)}_{\quad{\cal X}}\widehat{\cal H}_{ab}$ as
\begin{equation}
  \label{eq:(1,1)-order-metric-decomp}
  {}^{(1,1)}_{\quad{\cal X}}\widehat{\cal H}_{ab}
  =: {}^{(1,1)}_{\quad{\cal X}}{\cal H}_{ab}
  + \nabla_{a}{}^{(1,1)}_{\quad{\cal X}}X_{b}
  + \nabla_{b}{}^{(1,1)}_{\quad{\cal X}}X_{a},
\end{equation}
where ${}^{(1,1)}_{\quad{\cal X}}{\cal H}_{ab}$ is gauge invariant and
${}^{(1,1)}_{\quad{\cal X}}X_{a}$ transforms as 
\begin{equation}
  \label{eq:(1,1)-X-a-gauge-trans}
  {}^{(1,1)}_{\quad{\cal Y}}X^{a}
  = {}^{(1,1)}_{\quad{\cal X}}X^{a}
  + \xi_{(1,1)}^{a}
  + \frac{1}{2}[\xi_{(0,1)},{}^{(1,0)}_{\quad{\cal X}}X]^{a} 
  + \frac{1}{2}[\xi_{(1,0)},{}^{(0,1)}_{\quad{\cal X}}X]^{a}.
\end{equation}

%************************************************

%%%%%%%%%%%%%%%%%%%%%%%%%%%%%%%%%%%%%%%%%%%%%%%%%%%%%%%%%%%%%%%%%%%%%
\subsubsection{Third order metric perturbation}

%************************************************

Finally, we give the definitions of the gauge invariant
variables for third order metric perturbations.
The procedure to define gauge invariant variables is completely 
parallel to that in the case of second order perturbations.
First, we define the variables 
${}^{(p,q)}\widehat{\cal H}_{ab}$ at each order in such a way
that their gauge transformation rules have the same form as
those for the linear perturbations. 
If we define the variable ${}^{(p,q)}\widehat{\cal H}_{ab}$, we can
extract the gauge invariant part of the higher order perturbation
theory from this variable using the procedure employed for the
linear perturbations of the metric.

%************************************************\\

The non-trivial point of this procedure lies only in defining
the variables ${}^{(p,q)}\widehat{\cal H}_{ab}$.
The remaining part is trivial, due to the assumption that there
exists a procedure for the linear perturbation theory. 
The definitions of the variables ${}^{(p,q)}\widehat{\cal H}_{ab}$ of
$O(\lambda^{3})$, $O(\lambda^{2}\epsilon)$, $O(\lambda\epsilon^{2})$
and $O(\epsilon^{3})$ are as follows:
\begin{eqnarray}
  {}^{(3,0)}_{\quad{\cal X}}\widehat{\cal H}_{ab}
  &:=& 
  {}^{(3,0)}_{\quad{\cal X}}h_{ab}
  - 3 {\pounds}_{{}^{(1,0)}_{\quad{\cal X}}X} 
      {}^{(2,0)}_{\quad{\cal X}}h_{ab}
  - 3 {\pounds}_{{}^{(2,0)}_{\quad{\cal X}}X} 
      {}^{(1,0)}_{\quad{\cal X}}h_{ab}
  + 3 {\pounds}_{{}^{(1,0)}_{\quad{\cal X}}X}^{2} 
      {}^{(1,0)}_{\quad{\cal X}}h_{ab}
  \nonumber\\
  && 
  + \left\{ 
    3 {\pounds}_{{}^{(1,0)}_{\quad{\cal X}}X} 
      {\pounds}_{{}^{(2,0)}_{\quad{\cal X}}X}
    - {\pounds}_{{}^{(1,0)}_{\quad{\cal X}}X}^{3}
  \right\} {}^{(0)}g_{ab},
  \label{eq:widehat-H-ab-in-gauge-X-def-3.0}
  \\
  {}^{(2,1)}_{\quad{\cal X}}\widehat{\cal H}_{ab}
  &:=&
  {}^{(2,1)}_{\quad{\cal X}}h_{ab}
  -2{\pounds}_{{}^{(1,0)}_{\quad{\cal X}}X}{}^{(1,1)}_{\quad{\cal X}}h_{ab}
  - {\pounds}_{{}^{(0,1)}_{\quad{\cal X}}X}{}^{(2,0)}_{\quad{\cal X}}h_{ab}
  \nonumber\\
  && 
  - \left(
    {\pounds}_{{}^{(2,0)}_{\quad{\cal X}}X}
    - {\pounds}_{{}^{(1,0)}_{\quad{\cal X}}X}^{2}
  \right) {}^{(0,1)}_{\quad{\cal X}}h_{ab}
  \nonumber\\
  && 
  -2 \left\{
    {\pounds}_{{}^{(1,1)}_{\quad{\cal X}}X}
    - \frac{1}{2}
    \left( 
        {\pounds}_{{}^{(1,0)}_{\quad{\cal X}}X} 
        {\pounds}_{{}^{(0,1)}_{\quad{\cal X}}X} 
      + {\pounds}_{{}^{(0,1)}_{\quad{\cal X}}X} 
        {\pounds}_{{}^{(1,0)}_{\quad{\cal X}}X} 
    \right)
  \right\}
  {}^{(1,0)}_{\quad{\cal X}}h_{ab}
  \nonumber\\
  && 
  +  \left(
      2 {\pounds}_{{}^{(1,0)}_{\quad{\cal X}}X} 
        {\pounds}_{{}^{(1,1)}_{\quad{\cal X}}X}
    +   {\pounds}_{{}^{(0,1)}_{\quad{\cal X}}X}
        {\pounds}_{{}^{(2,0)}_{\quad{\cal X}}X}
  \right.
  \nonumber\\
  && \quad\quad\quad
  \left.
    -   {\pounds}_{{}^{(1,0)}_{\quad{\cal X}}X}
        {\pounds}_{{}^{(0,1)}_{\quad{\cal X}}X}
        {\pounds}_{{}^{(1,0)}_{\quad{\cal X}}X}
   \right) {}^{(0)}g_{ab} , 
  \label{eq:widehat-H-ab-in-gauge-X-def-2.1}
  \\
  {}^{(1,2)}_{\quad{\cal X}}\widehat{\cal H}_{ab}
  &:=&
  {}^{(1,2)}_{\quad{\cal X}}h_{ab}
  -2{\pounds}_{{}^{(0,1)}_{\quad{\cal X}}X}{}^{(1,1)}_{\quad{\cal X}}h_{ab}
  - {\pounds}_{{}^{(1,0)}_{\quad{\cal X}}X}{}^{(0,2)}_{\quad{\cal X}}h_{ab}
  \nonumber\\
  && 
  - \left(
    {\pounds}_{{}^{(0,2)}_{\quad{\cal X}}X}
    - {\pounds}_{{}^{(0,1)}_{\quad{\cal X}}X}^{2}
  \right) {}^{(1,0)}_{\quad{\cal X}}h_{ab}
  \nonumber\\
  && 
  - 2 \left\{
    {\pounds}_{{}^{(1,1)}_{\quad{\cal X}}X}
  - \frac{1}{2}\left(
      {\pounds}_{{}^{(1,0)}_{\quad{\cal X}}X}
      {\pounds}_{{}^{(0,1)}_{\quad{\cal X}}X} 
    + {\pounds}_{{}^{(0,1)}_{\quad{\cal X}}X}
      {\pounds}_{{}^{(1,0)}_{\quad{\cal X}}X} 
  \right)
  \right\}
  {}^{(0,1)}_{\quad{\cal X}}h_{ab}
  \nonumber\\
  &&
  +  \left(
      2 {\pounds}_{{}^{(0,1)}_{\quad{\cal X}}X}
        {\pounds}_{{}^{(1,1)}_{\quad{\cal X}}X}
    +   {\pounds}_{{}^{(1,0)}_{\quad{\cal X}}X}
        {\pounds}_{{}^{(0,2)}_{\quad{\cal X}}X}
  \right.
  \nonumber\\
  && \quad\quad\quad
  \left.
    -   {\pounds}_{{}^{(0,1)}_{\quad{\cal X}}X}
        {\pounds}_{{}^{(1,0)}_{\quad{\cal X}}X}
        {\pounds}_{{}^{(0,1)}_{\quad{\cal X}}X}
  \right) {}^{(0)}g_{ab}, 
  \label{eq:widehat-H-ab-in-gauge-X-def-1.2}
  \\
  {}^{(0,3)}_{\quad{\cal X}}\widehat{\cal H}_{ab}
  &:=& 
  {}^{(0,3)}_{\quad{\cal X}}h_{ab}
  -3{\pounds}_{{}^{(0,1)}_{\quad{\cal X}}X}{}^{(0,2)}_{\quad{\cal X}}h_{ab}
  -3{\pounds}_{{}^{(0,2)}_{\quad{\cal X}}X}{}^{(0,1)}_{\quad{\cal X}}h_{ab}
  +3{\pounds}_{{}^{(0,1)}_{\quad{\cal X}}X}^{2}{}^{(0,1)}_{\quad{\cal X}}h_{ab}
  \nonumber\\
  && \quad\quad
  + \left\{ 3 {\pounds}_{{}^{(0,1)}_{\quad{\cal X}}X} 
    {\pounds}_{{}^{(0,2)}_{\quad{\cal X}}X}
    - {\pounds}_{{}^{(0,1)}_{\quad{\cal X}}X}^{3}
  \right\} {}^{(0)}g_{ab}
  \label{eq:widehat-H-ab-in-gauge-X-def-0.3}
\end{eqnarray}
with the gauge ${\cal X}$.
The variables ${}^{(p,q)}_{\quad{\cal Y}}\widehat{\cal H}_{ab}$
for $p+q=3$ are defined for the gauge ${\cal Y}$ in same
forms as Eqs.~
(\ref{eq:widehat-H-ab-in-gauge-X-def-3.0})--(\ref{eq:widehat-H-ab-in-gauge-X-def-0.3})
with the replacement ${\cal X}\rightarrow{\cal Y}$.
Using the gauge transformation rules
(\ref{eq:metric-gauge-trans-1.0})--(\ref{eq:metric-gauge-trans-0.3}),
(\ref{eq:(1,0)-and-(0,1)-X-a-gauge-trans}),
(\ref{eq:(0,2)-X-a-gauge-trans}) and (\ref{eq:(1,1)-X-a-gauge-trans}), 
we can easily check that the variable 
${}^{(p,q)}_{\quad{\cal X}}\widehat{\cal H}_{ab}$ ($p+q=3$) is 
transformed as 
\begin{eqnarray}
  \label{eq:widehat-metric-gauge-trans-p+q=3}
  {}^{(p,q)}_{\quad{\cal Y}}\widehat{\cal H}_{ab}
  - {}^{(p,q)}_{\quad{\cal X}}\widehat{\cal H}_{ab}
  = {\pounds}_{\sigma_{(p,q)}} {}^{(0)}g_{ab}, 
\end{eqnarray}
where the vector fields $\sigma_{(p,q)}^{a}$ are defined by 
\begin{eqnarray}
  \sigma_{(3,0)}^{a} &:=& \xi_{(3,0)}^{a} 
  - 3 \left[
    \xi_{(2,0)} + {}^{(2,0)}_{\quad{\cal X}}X, \xi_{(1,0)}
  \right]^{a}
  \nonumber\\
  && \quad\quad\quad\quad
  + \left[
    2 \xi_{(1,0)} + {}^{(1,0)}_{\quad{\cal X}}X,
    \left[
      \xi_{(1,0)},{}^{(1,0)}_{\quad{\cal X}}X
    \right]
  \right]^{a},
  \\
  \sigma_{(2,1)}^{a} &:=& \xi_{(2,1)}^{a}
  + \left[
    \left[
      {}^{(0,1)}_{\quad{\cal X}}X,\xi_{(1,0)}
    \right]
    - 2{}^{(1,1)}_{\quad{\cal X}}X-2\xi_{(1,1)},\xi_{(1,0)}
  \right]^{a}
  \nonumber\\
  && \quad\quad\quad\quad
  + \left[
    \left[
      {}^{(1,0)}_{\quad{\cal X}}X,\xi_{(1,0)}
    \right]
    -{}^{(2,0)}_{\quad{\cal X}}X-\xi_{(2,0)},\xi_{(0,1)}
  \right]^{a} 
  \nonumber\\
  && \quad\quad\quad\quad
  + \left[
    {}^{(1,0)}_{\quad{\cal X}}X,
    \left[
      \xi_{(1,0)},{}^{(0,1)}_{\quad{\cal X}}X
    \right]
  \right]^{a},
  \\
  \sigma_{(1,2)}^{a} &:=& \xi_{(1,2)}^{a}
  + \left[
    \left[
      {}^{(1,0)}_{\quad{\cal X}}X,\xi_{(0,1)}
    \right]
    - 2 {}^{(1,1)}_{\quad{\cal X}}X - 2 \xi_{(1,1)},\xi_{(0,1)}
  \right]^{a}
  \nonumber\\
  && \quad\quad\quad\quad
  + \left[
    \left[
      {}^{(0,1)}_{\quad{\cal X}}X,\xi_{(1,0)}
    \right]
    - {}^{(0,2)}_{\quad{\cal X}}X - \xi_{(0,2)},\xi_{(1,0)}
  \right]^{a}
  \nonumber\\
  && \quad\quad\quad\quad
  + \left[
    {}^{(0,1)}_{\quad{\cal X}}X,
    \left[
      \xi_{(0,1)},{}^{(1,0)}_{\quad{\cal X}}X
    \right]
  \right]^{a},
  \\
  \sigma_{(0,3)}^{a} &:=& \xi_{(0,3)}^{a}
  - 3 \left[
    \xi_{(0,2)} + {}^{(0,2)}_{\quad{\cal X}}X
    ,\xi_{(0,1)}
  \right]^{a}
  \nonumber\\
  && \quad\quad\quad\quad
  + \left[
    2 \xi_{(0,1)} + {}^{(0,1)}_{\quad{\cal X}}X,
    \left[
      \xi_{(0,1)},{}^{(0,1)}_{\quad{\cal X}}X
    \right]
  \right]^{a}.
\end{eqnarray}

The gauge transformations given in 
(\ref{eq:widehat-metric-gauge-trans-p+q=3}) imply that we can
decompose the variables ${}^{(p,q)}_{\quad{\cal X}}\widehat{\cal H}_{ab}$
as
\begin{equation}
  {}^{(p,q)}_{\quad{\cal X}}\widehat{\cal H}_{ab} :=
  {}^{(p,q)}_{\quad{\cal X}}{\cal H}_{ab}
  + \nabla_{a}{}^{(p,q)}_{\quad{\cal X}}X_{b}
  + \nabla_{b}{}^{(p,q)}_{\quad{\cal X}}X_{a},
\end{equation}
using the procedure to find the gauge invariant variables for
linear order perturbations, where
${}^{(p,q)}_{\quad{\cal X}}{\cal H}_{ab}$ is gauge invariant and 
${}^{(p,q)}_{\quad{\cal X}}X_{a}$ at each order is transformed as 
\begin{eqnarray}
  \label{eq:(3,0)-X-a-gauge-trans}
  {}^{(3,0)}_{\quad{\cal Y}}X^{a} - {}^{(3,0)}_{\quad{\cal X}}X^{a}
  &=&
  \xi_{(3,0)}^{a} 
  - 3 \left[
    \xi_{(2,0)}
    + {}^{(2,0)}_{\quad{\cal X}}X
    ,\xi_{(1,0)}
  \right]^{a}
  \nonumber\\
  &&
  + \left[
    2 \xi_{(1,0)} + {}^{(1,0)}_{\quad{\cal X}}X,
    \left[
      \xi_{(1,0)},{}^{(1,0)}_{\quad{\cal X}}X
    \right]
  \right]^{a},
  \\
  \label{eq:(2,1)-X-a-gauge-trans}
  {}^{(2,1)}_{\quad{\cal Y}}X^{a} - {}^{(2,1)}_{\quad{\cal X}}X^{a}
  &=& \xi_{(2,1)}^{a}
  + \left[
    \left[
      {}^{(0,1)}_{\quad{\cal X}}X,\xi_{(1,0)}
    \right]
    - 2{}^{(1,1)}_{\quad{\cal X}}X-2\xi_{(1,1)},\xi_{(1,0)}
  \right]^{a}
  \nonumber\\
  && 
  + \left[
    \left[
      {}^{(1,0)}_{\quad{\cal X}}X,\xi_{(1,0)}
    \right]
    -{}^{(1,1)}_{\quad{\cal X}}X-\xi_{(1,1)},\xi_{(0,1)}
  \right]^{a}
  \nonumber\\
  && 
  + \left[
    {}^{(1,0)}_{\quad{\cal X}}X,
    \left[
      \xi_{(1,0)},{}^{(0,1)}_{\quad{\cal X}}X
    \right]
  \right]^{a},
  \\
  \label{eq:(1,2)-X-a-gauge-trans}
  {}^{(1,2)}_{\quad{\cal Y}}X^{a} - {}^{(1,2)}_{\quad{\cal X}}X^{a}
  &=& \xi_{(1,2)}^{a}
  + \left[
    \left[
      {}^{(1,0)}_{\quad{\cal X}}X,\xi_{(0,1)}
    \right]
    - 2 {}^{(1,1)}_{\quad{\cal X}}X - 2 \xi_{(1,1)},\xi_{(0,1)}
  \right]^{a}
  \nonumber\\
  && 
  + \left[
    \left[
      {}^{(0,1)}_{\quad{\cal X}}X,\xi_{(1,0)}
    \right]
    - {}^{(0,2)}_{\quad{\cal X}}X - \xi_{(0,2)},\xi_{(1,0)}
  \right]^{a}
  \nonumber\\
  && 
  + \left[
    {}^{(0,1)}_{\quad{\cal X}}X,
    \left[
      \xi_{(0,1)},{}^{(1,0)}_{\quad{\cal X}}X
    \right]
  \right]^{a},
  \\
  \label{eq:(0,3)-X-a-gauge-trans}
  {}^{(0,3)}_{\quad{\cal Y}}X^{a} - {}^{(0,3)}_{\quad{\cal X}}X^{a}
  &=& \xi_{(0,3)}^{a}
  - 3 \left[
    \xi_{(0,2)} + {}^{(0,2)}_{\quad{\cal X}}X
    ,\xi_{(0,1)}
  \right]^{a}
  \nonumber\\
  &&
  + 2 \left[
    2 \xi_{(0,1)} + {}^{(0,1)}_{\quad{\cal X}}X,
    \left[
      \xi_{(0,1)},{}^{(0,1)}_{\quad{\cal X}}X
    \right]
  \right]^{a}.
\end{eqnarray}

%************************************************

Thus, we have found recursively the gauge invariant variables
for higher order metric perturbations up to third order.

%%%%%%%%%%%%%%%%%%%%%%%%%%%%%%%%%%%%%%%%%%%%%%%%%%%%%%%%%%%%%%%%%%%%%
\subsection{Gauge invariant variables for matter perturbations}
\label{sec:matters}
%%%%%%%%%%%%%%%%%%%%%%%%%%%%%%%%%%%%%%%%%%%%%%%%%%%%%%%%%%%%%%%%%%%%%

As shown above, we can find the gauge invariant variables of
higher order metric perturbations.
To do so, we have decomposed the metric perturbations at
each order into the gauge invariant variables 
${}^{(p,q)}_{\quad{\cal X}}{\cal H}_{ab}$ and the gauge variant
vector variables ${}^{(p,q)}_{\quad{\cal X}}X_{a}$. 
The gauge variant parts, ${}^{(p,q)}_{\quad{\cal X}}X_{a}$ of the
$O(\lambda^{p}\epsilon^{q})$ metric perturbations are irrelevant
as physical metric perturbations.
However, using these gauge variant parts,
we can define the gauge invariant variables for the physical
fields, other than the metric.

%************************************************\\

Here, we give explicit definitions of gauge invariant variables
for perturbations of an arbitrary tensor field $Q$ up to third
order: 
\begin{eqnarray}
  \label{eq:matter-gauge-inv-def-1.0} 
  \delta^{(1,0)}_{{\cal X}}{\cal Q} &:=&
  \delta^{(1,0)}_{{\cal X}}Q 
  - {\pounds}_{{}^{(1,0)}_{\quad{\cal X}}X}Q_{0}
  , \\ 
  \label{eq:matter-gauge-inv-def-0.1} 
  \delta^{(0,1)}_{\cal X}{\cal Q} &:=&
  \delta^{(0,1)}_{\cal X}Q 
  - {\pounds}_{{}^{(0,1)}_{\quad{\cal X}}X}Q_{0}
  , \\
  \label{eq:matter-gauge-inv-def-2.0} 
  \delta^{(2,0)}_{\cal X}{\cal Q} &:=&
  \delta^{(2,0)}_{\cal X}Q 
  - 2 {\pounds}_{{}^{(1,0)}_{\quad{\cal X}}X} \delta^{(1,0)}_{\cal X}Q 
  - \left\{
    {\pounds}_{{}^{(2,0)}_{\quad{\cal X}}X}
    -{\pounds}_{{}^{(1,0)}_{\quad{\cal X}}X}^{2}
  \right\} Q_{0}
  , \\
  \label{eq:matter-gauge-inv-def-1.1} 
  \delta^{(1,1)}_{\cal X}{\cal Q} &:=&
  \delta^{(1,1)}_{\cal X}Q 
  - {\pounds}_{{}^{(1,0)}_{\quad{\cal X}}X} \delta^{(0,1)}_{\cal X}Q 
  - {\pounds}_{{}^{(0,1)}_{\quad{\cal X}}X} \delta^{(1,0)}_{\cal X}Q 
  \nonumber\\
  && \quad
  - \left\{{\pounds}_{{}^{(1,1)}_{\quad{\cal X}}X} 
    - \frac{1}{2} {\pounds}_{{}^{(1,0)}_{\quad{\cal X}}X} 
                  {\pounds}_{{}^{(0,1)}_{\quad{\cal X}}X} 
    - \frac{1}{2} {\pounds}_{{}^{(0,1)}_{\quad{\cal X}}X} 
                  {\pounds}_{{}^{(1,0)}_{\quad{\cal X}}X}
  \right\} Q_{0}
  ,\\
  \label{eq:matter-gauge-inv-def-0.2} 
  \delta^{(0,2)}_{\cal X}{\cal Q} &=&
  \delta^{(0,2)}_{\cal X}Q
  - 2 {\pounds}_{{}^{(0,1)}_{\quad{\cal X}}X} \delta^{(0,1)}_{\cal X}Q 
  -\left\{
    {\pounds}_{{}^{(0,2)}_{\quad{\cal X}}X}
    - {\pounds}_{{}^{(0,2)}_{\quad{\cal X}}X}^{2}
  \right\} Q_{0}
  ,\\
  \label{eq:matter-gauge-inv-def-3.0} 
  \delta^{(3,0)}_{\cal X}{\cal Q} &:=&
  \delta^{(3,0)}_{\cal X}Q 
  - 3 {\pounds}_{{}^{(1,0)}_{\quad{\cal X}}X} \delta^{(2,0)}_{\cal X}Q 
  - 3 \left\{ 
    {\pounds}_{{}^{(2,0)}_{\quad{\cal X}}X} 
    - {\pounds}_{{}^{(1,0)}_{\quad{\cal X}}X}^{2} 
  \right\}
  \delta^{(1,0)}_{\cal X}Q 
  \nonumber\\
  && \quad
  - \left\{ 
    {\pounds}_{{}^{(3,0)}_{\quad{\cal X}}X} 
    - 3 {\pounds}_{{}^{(1,0)}_{\quad{\cal X}}X}
        {\pounds}_{{}^{(2,0)}_{\quad{\cal X}}X} 
    + {\pounds}_{{}^{(1,0)}_{\quad{\cal X}}X}^{3} 
  \right\} Q_{0}
  , \\ 
  \label{eq:matter-gauge-inv-def-2.1}
  \delta^{(2,1)}_{\cal X}{\cal Q} &:=& \delta^{(2,1)}_{\cal X}Q 
  - 2 {\pounds}_{{}^{(1,0)}_{\quad{\cal X}}X} \delta^{(1,1)}_{\cal X}Q 
  - {\pounds}_{{}^{(1,0)}_{\quad{\cal X}}X} \delta^{(2,0)}_{\cal X}Q 
  \nonumber\\
  && \quad 
  - \left\{ 
    {\pounds}_{{}^{(2,0)}_{\quad{\cal X}}X} 
    - {\pounds}_{{}^{(1,0)}_{\quad{\cal X}}X}^{2} 
  \right\} 
  \delta^{(0,1)}_{\cal X}Q 
  \nonumber\\
  && \quad 
  - 2 \left\{ {\pounds}_{{}^{(1,1)}_{\quad{\cal X}}X} 
    - \frac{1}{2} {\pounds}_{{}^{(1,0)}_{\quad{\cal X}}X} 
                  {\pounds}_{{}^{(0,1)}_{\quad{\cal X}}X} 
    - \frac{1}{2} {\pounds}_{{}^{(0,1)}_{\quad{\cal X}}X} 
                  {\pounds}_{{}^{(1,0)}_{\quad{\cal X}}X}
  \right\} \delta^{(1,0)}_{\cal X}Q 
  \nonumber\\
  && \quad 
  - \left\{ 
        {\pounds}_{{}^{(2,1)}_{\quad{\cal X}}X} 
    - 2 {\pounds}_{{}^{(1,0)}_{\quad{\cal X}}X}
        {\pounds}_{{}^{(1,1)}_{\quad{\cal X}}X}
    -   {\pounds}_{{}^{(0,1)}_{\quad{\cal X}}X} 
        {\pounds}_{{}^{(2,0)}_{\quad{\cal X}}X} 
  \right.
  \nonumber\\
  && \quad\quad\quad
  \left.
    +   {\pounds}_{{}^{(1,0)}_{\quad{\cal X}}X} 
        {\pounds}_{{}^{(0,1)}_{\quad{\cal X}}X}
        {\pounds}_{{}^{(1,0)}_{\quad{\cal X}}X}
  \right\} Q_{0}
  , \\ 
  \label{eq:matter-gauge-inv-def-1.2}
  \delta^{(1,2)}_{\cal X}{\cal Q} &:=& 
  \delta^{(1,2)}_{\cal X}Q
  - 2 {\pounds}_{{}^{(0,1)}_{\quad{\cal X}}X} \delta^{(1,1)}_{\cal X}Q 
  -   {\pounds}_{{}^{(1,0)}_{\quad{\cal X}}X} \delta^{(0,2)}_{\cal X}Q 
  \nonumber\\
  && \quad 
  - \left\{ 
      {\pounds}_{{}^{(0,2)}_{\quad{\cal X}}X}
    - {\pounds}_{{}^{(0,1)}_{\quad{\cal X}}X}^{2} 
  \right\} 
  \delta^{(1,0)}_{\cal X}Q 
  \nonumber\\
  && \quad 
  - 2 \left\{ 
      {\pounds}_{{}^{(1,1)}_{\quad{\cal X}}X}
    - \frac{1}{2} {\pounds}_{{}^{(1,0)}_{\quad{\cal X}}X}
                  {\pounds}_{{}^{(0,1)}_{\quad{\cal X}}X}
    - \frac{1}{2} {\pounds}_{{}^{(0,1)}_{\quad{\cal X}}X}
                  {\pounds}_{{}^{(1,0)}_{\quad{\cal X}}X}
  \right\} 
  \delta^{(0,1)}_{\cal X}Q 
  \nonumber\\
  && \quad 
  - \left\{ 
        {\pounds}_{{}^{(1,2)}_{\quad{\cal X}}X}
    - 2 {\pounds}_{{}^{(0,1)}_{\quad{\cal X}}X}
        {\pounds}_{{}^{(1,1)}_{\quad{\cal X}}X}
    -   {\pounds}_{{}^{(1,0)}_{\quad{\cal X}}X} 
        {\pounds}_{{}^{(0,2)}_{\quad{\cal X}}X} 
  \right.
  \nonumber\\
  && \quad\quad\quad
  \left.
    +   {\pounds}_{{}^{(0,1)}_{\quad{\cal X}}X}
        {\pounds}_{{}^{(1,0)}_{\quad{\cal X}}X}
        {\pounds}_{{}^{(0,1)}_{\quad{\cal X}}X}
  \right\} Q_{0}
  , \\ 
  \label{eq:matter-gauge-inv-def-0.3}
  \delta^{(0,3)}_{\cal X}{\cal Q} &:=& \delta^{(0,3)}_{\cal X}Q
  - 3 {\pounds}_{{}^{(0,1)}_{\quad{\cal X}}X} \delta^{(0,2)}_{\cal X}Q 
  - 3 \left\{ 
      {\pounds}_{{}^{(0,2)}_{\quad{\cal X}}X}
    - {\pounds}_{{}^{(0,1)}_{\quad{\cal X}}X}^{2}
  \right\}
  \delta^{(0,1)}_{\cal X}Q 
  \nonumber\\
  && \quad
  - \left\{ 
        {\pounds}_{{}^{(0,3)}_{\quad{\cal X}}X}
    - 3 {\pounds}_{{}^{(0,1)}_{\quad{\cal X}}X}
        {\pounds}_{{}^{(0,2)}_{\quad{\cal X}}X}
    +   {\pounds}_{{}^{(0,1)}_{\quad{\cal X}}X}^{3}
  \right\} Q_{0}.
\end{eqnarray}
Straightforward calculations using the gauge transformation rules
(\ref{eq:Bruni-47})--(\ref{eq:Bruni-55}),
(\ref{eq:(1,0)-and-(0,1)-X-a-gauge-trans}),
(\ref{eq:(2,0)-X-a-gauge-trans}), (\ref{eq:(0,2)-X-a-gauge-trans}),
(\ref{eq:(1,1)-X-a-gauge-trans}) and
(\ref{eq:(3,0)-X-a-gauge-trans})--(\ref{eq:(0,3)-X-a-gauge-trans})
show that these variables $\delta^{(p,q)}_{\cal X}{\cal Q}$ are gauge
invariant.

%************************************************\\

%%%%%%%%%%%%%%%%%%%%%%%%%%%%%%%%%%%%%%%%%%%%%%%%%%%%%%%%%%%%%%%%%%%%%
\section{Summary and Discussions}
\label{sec:Summary-Discussions}
%%%%%%%%%%%%%%%%%%%%%%%%%%%%%%%%%%%%%%%%%%%%%%%%%%%%%%%%%%%%%%%%%%%%%

%************************************************

In this paper, we have presented the procedure to find gauge invariant
variables of two-parameter nonlinear metric and matter perturbations. 
To describe this procedure, we have assumed the existence of the
analogous procedure for linear perturbations. 
As emphasized in the main text, the decomposition of the linear
perturbation of the metric into gauge invariant and gauge
variant parts is non-trivial.
It depends crucially on the background spacetime.
However, if we assume the existence of this procedure, we can
always define the gauge invariant variables for higher order
metric and matter perturbations.

%************************************************

The procedure presented above is summarized as follows.
Suppose that gauge transformation rules for a $(p,q)$-order
metric perturbation ${}^{(p,q)}h_{ab}$ are given by 
\begin{equation}
  {}^{(p,q)}_{\quad{\cal Y}}h_{ab} - {}^{(p,q)}_{\quad{\cal X}}h_{ab} 
  = F[\xi_{(p,q)}^{a},\xi_{(i,j)}^{a},{}^{(i,j)}_{\quad{\cal X}}h_{ab}], 
\end{equation}
where $F$ is a function determined by the gauge transformation rule
${\cal X}\rightarrow{\cal Y}$ of the $(p,q)$-order metric perturbation
${}^{(p,q)}h_{ab}$, and $i$ and $j$ are integers that satisfy the
conditions $i\leq p$, $j\leq q$ and $i+j\neq p+q$.
Further, suppose that for any $i$ and $j$, we have already
defined the tensor fields 
${}^{(i,j)}_{\quad{\cal X}}{\cal H}_{ab}$ and the vector fields
${}^{(i,j)}_{\quad{\cal X}}X^{a}$ as in the main text. 
Then, defining the variable 
${}^{(p,q)}_{\quad{\cal X}}\widehat{\cal H}_{ab}$ 
for the $O(\lambda^{p}\epsilon^{q})$ metric perturbation by  
\begin{equation}
  {}^{(p,q)}_{\quad{\cal X}}\widehat{\cal H}_{ab}
  := {}^{(p,q)}_{\quad{\cal X}}h_{ab} 
  + F[\xi_{(p,q)}^{a},-{}^{(i,j)}_{\quad{\cal X}}X^{a}
  ,{}^{(i,j)}_{\quad{\cal X}}h_{ab}], 
\end{equation}
there exists a vector field $\sigma_{(p,q)}^{a}$ such that 
the variable ${}^{(p,q)}_{\quad{\cal X}}\widehat{\cal H}_{ab}$
transforms as 
\begin{equation}
  {}^{(p,q)}_{\quad{\cal Y}}\widehat{\cal H}_{ab}
  - {}^{(p,q)}_{\quad{\cal X}}\widehat{\cal H}_{ab} 
  =  {\pounds}_{\sigma_{(p,q)}} {}^{(0)}g_{ab}
\end{equation}
under the gauge transformation ${\cal X}\rightarrow{\cal Y}$.
Then, using the same procedure as for linear perturbations, 
we can decompose ${}^{(p,q)}_{\quad{\cal X}}\widehat{\cal H}_{ab}$
as
\begin{equation}
  {}^{(p,q)}_{\quad{\cal X}}\widehat{\cal H}_{ab}
  =: {}^{(p,q)}_{\quad{\cal X}}{\cal H}_{ab}
  + \nabla_{a}{}^{(p,q)}_{\quad{\cal X}}X_{b}
  + \nabla_{b}{}^{(p,q)}_{\quad{\cal X}}X_{a},
\end{equation}
where the variables ${}^{(p,q)}_{\quad{\cal X}}{\cal H}_{ab}$ and 
${}^{(p,q)}_{\quad{\cal X}}X_{a}$ are the gauge invariant and
gauge variant parts of the $O(\lambda^{p}\epsilon^{q})$ metric
perturbation, respectively. 
Because the gauge transformation rule for the matter perturbations
$\delta^{(p,q)}Q$ is given by 
\begin{equation}
  \delta^{(p,q)}_{\cal Y}Q - \delta^{(p,q)}_{\cal X}Q 
  = F[\xi_{(p,q)}^{a},\xi_{(i,j)}^{a},\delta^{(i,j)}_{\cal X}Q], 
\end{equation}
the corresponding gauge invariant variables 
$\delta^{(p,q)}_{\cal X}{\cal Q}$ are defined by 
\begin{equation}
  \delta^{(p,q)}_{\cal X}{\cal Q}
  = 
  \delta^{(p,q)}_{\cal X}Q
  +
  F[-{}^{(p,q)}_{\quad{\cal X}}X^{a}
  ,-{}^{(i,j)}_{\quad{\cal X}}X^{a},\delta^{(i,j)}_{\cal X}Q].
\end{equation}

%************************************************

A procedure similar to that presented here for second order
perturbations in the one-parameter case was previously obtained
by Campanelli and Lousto\cite{CAmpanelli-Lousto1999}.
They applied that procedure to second order perturbations of a
Kerr black hole.
We have confirmed their procedure up to third order in the
two-parameter case.
Though the gauge transformation rule for the perturbations of
arbitrary order is not yet known, we conjecture that the
procedure considered here is applicable to arbitrary order
perturbations. 
We also believe that this procedure can be confirmed by
induction, once the gauge transformation rule for an arbitrary
order is obtained.
We leave this for a future work.

%************************************************

In addition to the interesting mathematical framework, it is also
interesting to apply it to the astrophysical systems, such as
oscillating relativistic rotating stars.
Many astrophysical systems can be described well by perturbation
theory with two parameters.
One merit of applying the procedure presented here to such
systems is the gauge ambiguities are removed.
When we apply this procedure to oscillating stars, a careful
analysis is necessary to properly treat the boundary conditions at the 
surface of the star and the displacement of this surface.
Similar situations are considered in preveous papers coauthored
by the present author\cite{kouchan-string}.
In those papers, a particular gauge fixing is necessary to match the
perturbative solutions at the boundary of the surface of the matter
distribution when we construct global solutions and when we define the
perturbative displacement of the matter surface.
Similar problems also arise in the perturbative analysis
of spacetimes with boundaries, such as brane worlds\cite{Randall-Sundrum}.
In the investigation of these situations, the gauge
transformation rules derived in BGS2003 and the gauge invariant
variables defined here become powerful tools.
Because of their applicability to various situations, we also expect
that the techniques developed here will play a key role in
progress of theoretical physics.

%************************************************

%%%%%%%%%%%%%%%%%%%%%%%%%%%%%%%%%%%%%%%%%%%%%%%%%%%%%%%%%%%%%%%%%%%%%%
\section*{Acknowledgements}
%%%%%%%%%%%%%%%%%%%%%%%%%%%%%%%%%%%%%%%%%%%%%%%%%%%%%%%%%%%%%%%%%%%%%%
We would like to thank to Prof. Minoru Omote (Keio University) for his
continuous encouragement.

%%%%%%%%%%%%%%%%%%%%%%%%%%%%%%%%%%%%%%%%%%%%%%%%%%%%%%%%%%%%%%%%%%%%%%%
\appendix
%%%%%%%%%%%%%%%%%%%%%%%%%%%%%%%%%%%%%%%%%%%%%%%%%%%%%%%%%%%%%%%%%%%%%%%

%%%%%%%%%%%%%%%%%%%%%%%%%%%%%%%%%%%%%%%%%%%%%%%%%%%%%%%%%%%%%%%%%%%%%%%
\section{Equivalence with the Representation of Bruni et al.}
\label{sec:taylor-derivation}
%%%%%%%%%%%%%%%%%%%%%%%%%%%%%%%%%%%%%%%%%%%%%%%%%%%%%%%%%%%%%%%%%%%%%%%

In this appendix, we derive the representation of the
coefficients of the formal Taylor expansion
(\ref{eq:formal-Taylor-expansion}) of the pull-back of a
diffeomorphism in terms of the suitable derivative operators
${\cal L}_{(p,q)}$. 
As mentioned in the main text, for each of these operators,
there is a vector field such that the operators satisfy
Eq.~(\ref{eq:Lie-calL-rela}). 
The existence of these vector fields is guaranteed by the fact
that the operator ${\cal L}_{(p,q)}$ of each order satisfies
the Leibniz rule.
From this fact, the definitions of the derivative operators
(\ref{eq:Bruni-14})--(\ref{eq:Bruni-22}) are obtained. 
As shown in the Appendix in BGS2003, the representation of the 
Taylor expansion of $\Phi^{*}_{\lambda,\epsilon}f$ for an
arbitrary function $f$ can be extended to that for an arbitrary
tensor field $Q$ on ${\cal M}$. 
Therefore, in this appendix, we only consider the Taylor expansion
of the pull-back $\Phi^{*}_{\lambda,\epsilon}f$ for an arbitrary
scalar function $f\in {\cal F}({\cal M})$:
\begin{equation}
  \Phi^{*}_{\lambda,\epsilon} f = \sum^{\infty}_{k,k'=0}
  \frac{\lambda^{k}\epsilon^{k'}}{k!k'!} 
  \left\{
    \frac{\partial^{k+k'}}{\partial\lambda^{k}\partial\epsilon^{k'}} 
    \Phi^{*}_{\lambda,\epsilon} f
  \right\}_{\lambda=\epsilon=0},
  \label{eq:formal-Taylor-expansion-scalar-function}
\end{equation}
where ${\cal F}({\cal M})$ denotes the algebra of $C^{\infty}$
functions on ${\cal M}$.

%************************************************

Although the operators $\partial/\partial\lambda$ and
$\partial/\partial\epsilon$ in the bracket
$\{*\}_{\lambda=\epsilon=0}$ of
Eq.~(\ref{eq:formal-Taylor-expansion-scalar-function}) are simply 
symbolic notation, we stipulate the properties
\begin{eqnarray}
  \left\{
    \frac{\partial^{n+1}}{\partial\lambda^{n+1}}\Phi_{\lambda,\epsilon}^{*}f
  \right\}_{\lambda=\epsilon=0}
  &=& 
  \left\{
    \frac{\partial}{\partial\lambda}
    \left(
      \frac{\partial^{n}}{\partial\lambda^{n}}
      \Phi_{\lambda,\epsilon}^{*}f
    \right)
  \right\}_{\lambda=\epsilon=0} 
  \nonumber\\
  &=& 
  \left\{
      \frac{\partial^{n}}{\partial\lambda^{n}}
    \left(
      \frac{\partial}{\partial\lambda}
      \Phi_{\lambda,\epsilon}^{*}f
    \right)
  \right\}_{\lambda=\epsilon=0}, 
  \label{eq:imposing-properties-1}
  \\
  \left\{
    \frac{\partial^{n+1}}{\partial\epsilon^{n+1}}\Phi_{\lambda,\epsilon}^{*}f
  \right\}_{\lambda=\epsilon=0}
  &=& 
  \left\{
    \frac{\partial}{\partial\epsilon}
    \left(
      \frac{\partial^{n}}{\partial\epsilon^{n}}
      \Phi_{\lambda,\epsilon}^{*}f
    \right)
  \right\}_{\lambda=\epsilon=0} 
  \nonumber\\
  &=& 
  \left\{
      \frac{\partial^{n}}{\partial\epsilon^{n}}
    \left(
      \frac{\partial}{\partial\epsilon}
      \Phi_{\lambda,\epsilon}^{*}f
    \right)
  \right\}_{\lambda=\epsilon=0}, 
  \label{eq:imposing-properties-2}
  \\
  \left\{
    \frac{\partial^{2}}{\partial\lambda\partial\epsilon}
    \Phi_{\lambda,\epsilon}^{*}f
  \right\}_{\lambda=\epsilon=0}
  &=& 
  \left\{
    \frac{\partial}{\partial\lambda}
    \left(
      \frac{\partial}{\partial\epsilon}
      \Phi_{\lambda,\epsilon}^{*}f
    \right)
  \right\}_{\lambda=\epsilon=0}
  \nonumber\\
  &=& 
  \left\{
    \frac{\partial}{\partial\epsilon}
    \left(
      \frac{\partial}{\partial\lambda}
      \Phi_{\lambda,\epsilon}^{*}f
    \right)
  \right\}_{\lambda=\epsilon=0},
  \label{eq:imposing-properties-3}
  \\
  \left\{
    \frac{\partial}{\partial\lambda}(\Phi_{\lambda,\epsilon}^{*}f)^{2}
  \right\}_{\lambda=\epsilon=0}
  &=& 
  \left\{
    2 \Phi_{\lambda,\epsilon}^{*}f
    \frac{\partial}{\partial\lambda}
    \left(
      \Phi_{\lambda,\epsilon}^{*}f
    \right)
  \right\}_{\lambda=\epsilon=0}, 
  \label{eq:imposing-properties-4}
  \\
  \left\{
    \frac{\partial}{\partial\epsilon}(\Phi_{\lambda,\epsilon}^{*}f)^{2}
  \right\}_{\lambda=\epsilon=0}
  &=& 
  \left\{
    2 \Phi_{\lambda,\epsilon}^{*}f
    \frac{\partial}{\partial\epsilon}
    \left(
      \Phi_{\lambda,\epsilon}^{*}f
    \right)
  \right\}_{\lambda=\epsilon=0} 
  \label{eq:imposing-properties-5}
\end{eqnarray}
for $\forall f\in{\cal F}({\cal M})$, where $n$ is an arbitrary
finite integer. 
These properties imply that the operators
$\partial/\partial\lambda$ and $\partial/\partial\epsilon$ are
in fact not simply symbolic notation but indeed the usual
partial differential operators on $\MR^{2}$. 
We note that the properties (\ref{eq:imposing-properties-4}) and
(\ref{eq:imposing-properties-5}) are the Leibniz rules, which play
important roles when we derive the representation of the Taylor 
expansion (\ref{eq:formal-Taylor-expansion-scalar-function}) in terms
of suitable Lie derivatives.

%************************************************\\

We can easily show that the derivative operators 
${\cal L}_{(1,0)}$ and ${\cal L}_{(0,1)}$, defined by 
(\ref{eq:Bruni-14}) and (\ref{eq:Bruni-15}), respectively,
satisfy the Leibniz rule
\begin{equation}
  \label{eq:calL-Libnitz}
  {\cal L}_{(p,q)} f^{2} = 2 f {\cal L}_{(p,q)} f
\end{equation}
due to the properties (\ref{eq:imposing-properties-4}) and
(\ref{eq:imposing-properties-5}). 
In the higher order coefficients of the expansion
(\ref{eq:formal-Taylor-expansion-scalar-function}), the properties 
(\ref{eq:imposing-properties-4}), (\ref{eq:imposing-properties-5})
and (\ref{eq:calL-Libnitz}) lead to non-trivial combinations of
the linear operators.
In BGS2003, it is commented that the representations of the higher
order coefficients is not unique, and the following
combinations are derived:
\begin{eqnarray}
  \label{eq:Bruni-14-append}
  \left\{ 
    \frac{\partial}{\partial\lambda}\Phi^{*}_{\lambda,\epsilon}f
  \right\}_{\lambda=\epsilon=0} 
  &=& {\cal L}_{(1,0)}f , \\  
  \label{eq:Bruni-15-append}
  \left\{ 
    \frac{\partial}{\partial\epsilon}\Phi^{*}_{\lambda,\epsilon}f
  \right\}_{\lambda=\epsilon=0}
  &=& {\cal L}_{(0,1)}f , \\  
  \label{eq:Bruni-16-append}
  \left\{ 
    \frac{\partial^{2}}{\partial\lambda^{2}}\Phi^{*}_{\lambda,\epsilon}f
  \right\}_{\lambda=\epsilon=0} 
  &=& {\cal L}_{(2,0)}f + {\cal L}^{2}_{(1,0)}f, \\  
  \label{eq:Bruni-17-append}
  \left\{ 
    \frac{\partial^{2}}{\partial\lambda\partial\epsilon} 
    \Phi^{*}_{\lambda,\epsilon}f
  \right\}_{\lambda=\epsilon=0} 
  &=& 
  {\cal L}_{(1,1)}f
  + \left(
    \epsilon_{0}{\cal L}_{(1,0)}{\cal L}_{(0,1)} 
    + \epsilon_{1}{\cal L}_{(0,1)}{\cal L}_{(1,0)}
  \right)f ,\\ 
  \label{eq:Bruni-18-append}
  \left\{ 
    \frac{\partial^{2}}{\partial\epsilon^{2}}\Phi^{*}_{\lambda,\epsilon}f
  \right\}_{\lambda=\epsilon=0} 
  &=& 
  {\cal L}_{(0,2)}f + {\cal L}^{2}_{(0,1)}f, \\  
  \label{eq:Bruni-19-append}
  \left\{ 
    \frac{\partial^{3}}{\partial\lambda^{3}}\Phi^{*}_{\lambda,\epsilon}f
  \right\}_{\lambda=\epsilon=0} 
  &=& 
  {\cal L}_{(3,0)}f
  + 3 {\cal L}_{(1,0)} {\cal L}_{(2,0)}f 
  + {\cal L}^{3}_{(1,0)}f, \\  
  \left\{ 
    \frac{\partial^{3}}{\partial\lambda^{2}\partial\epsilon}
    \Phi^{*}_{\lambda,\epsilon}f
  \right\}_{\lambda=\epsilon=0} 
  &=& {\cal L}_{(2,1)}f
  + 2 {\cal L}_{(1,0)} {\cal L}_{(1,1)}f 
  \nonumber\\
  &&
  + {\cal L}_{(0,1)} {\cal L}_{(2,0)}f 
  + 2 \epsilon_{2} {\cal L}_{(1,0)} {\cal L}_{(0,1)} {\cal L}_{(1,0)} f 
  \nonumber\\
  &&
  \label{eq:Bruni-20-append}
  + (\epsilon_{1}-\epsilon_{2}) {\cal L}_{(0,1)} {\cal L}_{(1,0)}^{2} f
  + (\epsilon_{0}-\epsilon_{2}) {\cal L}^{2}_{(1,0)} {\cal L}_{(0,1)} f, \\
  \left\{ 
    \frac{\partial^{3}}{\partial\lambda\partial\epsilon^{2}}
    \Phi^{*}_{\lambda,\epsilon}f
  \right\}_{\lambda=\epsilon=0} 
  &=& {\cal L}_{(1,2)}f
  + 2 {\cal L}_{(0,1)} {\cal L}_{(1,1)}f 
  \nonumber\\
  &&
  + {\cal L}_{(1,0)} {\cal L}_{(0,2)}f 
  + 2 \epsilon_{3} {\cal L}_{(0,1)} {\cal L}_{(1,0)} {\cal L}_{(0,1)} f 
  \nonumber\\
  &&
  \label{eq:Bruni-21-append}
  + (\epsilon_{0}-\epsilon_{3}) {\cal L}_{(1,0)} {\cal L}_{(0,1)}^{2} f
  + (\epsilon_{1}-\epsilon_{3}) {\cal L}^{2}_{(0,1)} {\cal L}_{(1,0)} f, \\
  \left\{ 
    \frac{\partial^{3}}{\partial\epsilon^{3}}\Phi^{*}_{\lambda,\epsilon}f
  \right\}_{\lambda=\epsilon=0} 
  &=& {\cal L}_{(0,3)}f
  + 3 {\cal L}_{(0,1)} {\cal L}_{(0,2)}f 
  + {\cal L}^{3}_{(0,1)}f.
  \label{eq:Bruni-22-append}
\end{eqnarray}
Here, the parameters $\epsilon_{i}$ ($i=0,1,2,3$) are constants
that satisfy the condition 
\begin{equation}
  \label{eq:epsilon-1-0-constraint}
  \epsilon_{0} + \epsilon_{1} = 1.
\end{equation}
These parameters result from the fact that the representation of
the higher order coefficients in terms of the derivative
operators is not unique.

%************************************************\\

As emphasized in BGS2003, the parameters $\epsilon_{i}$ in
the representations
(\ref{eq:Bruni-14-append})--(\ref{eq:Bruni-22-append}) are not
essential.
Each operator ${\cal L}_{(p,q)}$ is regarded as the Lie
derivative with respect to the vector field $\xi_{(p,q)}^{a}$ as
mentioned above (see Eq.~(\ref{eq:Lie-calL-rela})). 
Once we obtain the representation of ${\cal L}_{(p,q)}$ in
terms of the Lie derivative, the parameters $\epsilon_{i}$ are
always eliminated through the replacement of the vector field.
As a result, we obtain the ``{\it canonical representation}''
(\ref{eq:two-parameter-Bruni-30-simpler}) in the main text.
Therefore, the ``canonical representation''
(\ref{eq:two-parameter-Bruni-30-simpler}) of the Taylor expansion with
two infinitesimal parameters is equivalent to that in BGS2003 as shown
below.

%************************************************\\

To show this, it is necessary to give an explicit derivation of the 
representations
(\ref{eq:Bruni-14-append})--(\ref{eq:Bruni-22-append}).
The derivations of these representations are done recursively from
lower order representations. 
Because the derivation of each representation is similar, it is enough
to present the explicit derivation of the representation of the
coefficients of $O(\lambda^{2}\epsilon)$, and we start from the
point where the lower order representations
(\ref{eq:Bruni-14-append})--(\ref{eq:Bruni-19-append}) are already
given. 
It is obvious that the coefficient of $O(\lambda^{2}\epsilon)$
in the expansion (\ref{eq:formal-Taylor-expansion-scalar-function})
includes a linear combination of the following terms:
\begin{eqnarray}
  && {\cal L}_{(0,1)}{\cal L}_{(2,0)}f, \quad
     {\cal L}_{(2,0)}{\cal L}_{(0,1)}f, \quad
     {\cal L}_{(1,0)}{\cal L}_{(1,1)}f, \quad
     {\cal L}_{(1,1)}{\cal L}_{(1,0)}f, \nonumber\\
  && {\cal L}_{(0,1)}{\cal L}_{(1,0)}{\cal L}_{(1,0)}f, \quad
     {\cal L}_{(1,0)}{\cal L}_{(0,1)}{\cal L}_{(1,0)}f, \quad
     {\cal L}_{(1,0)}{\cal L}_{(1,0)}{\cal L}_{(0,1)}f.
\end{eqnarray}
Then, we start from the following form of the derivative operator 
${\cal L}_{(2,1)}$: 
\begin{eqnarray}
  {\cal L}_{(2,1)}f &:=& \left\{ 
    \frac{\partial^{3}}{\partial\lambda^{2}\partial\epsilon}
    \Phi^{*}_{\lambda,\epsilon}f
  \right\}_{\lambda=\epsilon=0} \nonumber\\
  &&
  + p_{1} {\cal L}_{(0,1)}{\cal L}_{(2,0)}f
  + p_{2} {\cal L}_{(1,0)}{\cal L}_{(1,1)}f 
  \nonumber\\
  &&
  + p_{3} {\cal L}_{(0,1)}{\cal L}_{(1,0)}{\cal L}_{(1,0)}f
  + p_{4} {\cal L}_{(1,0)}{\cal L}_{(0,1)}{\cal L}_{(1,0)}f
  + p_{5} {\cal L}_{(1,0)}{\cal L}_{(1,0)}{\cal L}_{(0,1)}f
  \nonumber\\
  &&
  + p_{6} {\cal L}_{(2,0)}{\cal L}_{(0,1)}f
  + p_{7} {\cal L}_{(1,1)}{\cal L}_{(1,0)}f.
  \label{eq:(2,1)-starting-point}
\end{eqnarray}
Following the argument in BGS2003, we choose
\begin{equation}
  p_{6} = p_{7} = 0.
  \label{eq:(2,1)-p6-p7-vanishes}
\end{equation}
This choice is always possible.
The lower order derivative operators ${\cal L}_{(p,q)}$ are
regarded as the Lie derivative with appropriate generators
$\xi_{(p,q)}^{a}$ as in Eq.~(\ref{eq:Lie-calL-rela}) and the
first term on the second line and the first term on the fourth
line in the right-hand side of
Eq.~(\ref{eq:(2,1)-starting-point}) are given by 
\begin{eqnarray}
  && p_{1} {\cal L}_{(0,1)}{\cal L}_{(2,0)}f
  + p_{6} {\cal L}_{(2,0)}{\cal L}_{(0,1)}f
  \nonumber\\
  &=& (p_{1} + p_{6}) {\pounds}_{\xi_{(0,1)}}{\pounds}_{\xi_{(2,0)}}f
  + {\pounds}_{p_{6} [\xi_{(2,0)},\xi_{(0,1)}]}f.
\end{eqnarray}
Because, as we will show, 
${\cal L}_{(2,1)}={\pounds}_{\xi_{(2,1)}}$ even if 
$p_{6}\neq0$, we can always choose $p_{6}=0$ by making the
replacement
\begin{equation}
  \label{two-parameter-kouchan-comment-9.5}
  p_{1} + p_{6} \rightarrow p_{1}, \quad
  \xi_{(2,1)}^{a} + p_{6} [\xi_{(2,0)},\xi_{(0,1)}]^{a} \rightarrow 
  \xi_{(2,1)}^{a},
\end{equation}
without loss of generality.
The same argument can be applied to the parameter $p_{7}$.
Therefore, we may choose $p_{6}=p_{7}=0$, without loss of generality.
We note that similar arguments can be applied to the cases
of the parameter $p_{3}$, $p_{4}$ and $p_{5}$.
We can fix these parameters by using the replacement of the vector field
$\xi_{(2,1)}^{a}$. 
However, we do not proceed with this argument here, because we
have confirmed that the representation in BGS2003, i.e.
(\ref{eq:Bruni-14-append})--(\ref{eq:Bruni-22-append}) in this paper,
and Eqs.~(\ref{eq:Bruni-14})--(\ref{eq:Bruni-22}) are equivalent,
at least up to fourth order.

%************************************************\\

To guarantee the Leibniz rule for the derivative operator 
${\cal L}_{(2,1)}$, we first consider the operation 
\begin{equation}
  \left\{ 
    \frac{\partial^{3}}{\partial\lambda^{2}\partial\epsilon}
    (\Phi^{*}_{\lambda,\epsilon}f)^{2}
  \right\}_{\lambda=\epsilon=0} \quad \forall f\in{\cal F}({\cal M}).
  \label{eq:A.24}
\end{equation}
Using
Eqs.~(\ref{eq:imposing-properties-1})--(\ref{eq:imposing-properties-5}), 
we obtain 
\begin{eqnarray}
  \left\{ 
    \frac{\partial^{3}}{\partial\lambda^{2}\partial\epsilon}
    (\Phi^{*}_{\lambda,\epsilon}f)^{2}
  \right\}_{\lambda=\epsilon=0}
  &=& 
  2 \left\{ 
    \frac{\partial^{2}}{\partial\lambda^{2}}
    (\Phi^{*}_{\lambda,\epsilon}f)
  \right\}_{\lambda=\epsilon=0} 
  \left\{ 
    \frac{\partial}{\partial\epsilon}
    (\Phi^{*}_{\lambda,\epsilon}f)
  \right\}_{\lambda=\epsilon=0} 
  \nonumber\\
  && \quad
    + 4
  \left\{ 
    \frac{\partial}{\partial\lambda}
    (\Phi^{*}_{\lambda,\epsilon}f)
  \right\}_{\lambda=\epsilon=0} 
  \left\{ 
    \frac{\partial^{2}}{\partial\lambda\partial\epsilon}
    (\Phi^{*}_{\lambda,\epsilon}f)
  \right\}_{\lambda=\epsilon=0} 
  \nonumber\\
  && \quad
    + 2f
  \left\{ 
    \frac{\partial^{3}}{\partial\lambda^{2}\partial\epsilon}
    (\Phi^{*}_{\lambda,\epsilon}f)
  \right\}_{\lambda=\epsilon=0}.
  \label{eq:A.25}
\end{eqnarray}
Substituting Eqs.~(\ref{eq:Bruni-14-append})--(\ref{eq:Bruni-18-append}),
(\ref{eq:(2,1)-starting-point}) and (\ref{eq:(2,1)-p6-p7-vanishes})
into Eq.~(\ref{eq:A.25}), we obtain the representation of
Eq.~(\ref{eq:A.24}), which includes the term $2f{\cal L}_{(2,1)}f$.
On the other hand, a direct calculation from
Eq.~(\ref{eq:(2,1)-starting-point}) with 
Eqs.(\ref{eq:(2,1)-p6-p7-vanishes}) gives the representation of 
Eq.~(\ref{eq:A.24}) which includes the term ${\cal L}_{(2,1)}f^{2}$.
To obtain this second representation of Eq.~(\ref{eq:A.24}), we can
use the Leibniz rules for the derivative operators, except for 
${\cal L}_{(2,1)}$, because the Leibniz rules for the lower order
operators can be demonstrated by arguments similar to that given
here, and these can be confirmed before applying the arguments
for the operator ${\cal L}_{(2,1)}$.
From these two representations of Eq.~(\ref{eq:A.24}), we obtain 
\begin{eqnarray}
  {\cal L}_{(2,1)}(f^{2})
  - 2 f {\cal L}_{(2,1)} f
  &=&
    2 (p_{1} + 1) ({\cal L}_{(0,1)}f) ({\cal L}_{(2,0)}f)
  \nonumber\\
  &&\quad
  + 2 (p_{2} + 2) ({\cal L}_{(1,0)}f)({\cal L}_{(1,1)}f) 
  \nonumber\\
  &&\quad
  + 2 (p_{3} + p_{4} + p_{5} + 1) ({\cal L}_{(0,1)}f)({\cal L}_{(1,0)}^{2}f)
  \nonumber\\
  &&\quad
  + 2 (p_{4} + 2 p_{3} + 2 \epsilon_{1}) 
  ({\cal L}_{(0,1)}{\cal L}_{(1,0)}f)({\cal L}_{(1,0)}f)
  \nonumber\\
  &&\quad
  + 2 (p_{4} + 2 p_{5} + 2 \epsilon_{0})
  ({\cal L}_{(1,0)}{\cal L}_{(0,1)}f)({\cal L}_{(1,0)}f).
\end{eqnarray}
This shows that the derivative operator ${\cal L}_{(2,1)}$
satisfies the Leibniz rule 
${\cal L}_{(2,1)}(f^{2}) = 2 f {\cal L}_{(2,1)}f$ 
for an arbitrary function $f\in{\cal F}({\cal M})$ iff
\begin{eqnarray}
  p_{1} = - 1, \quad
  p_{2} = - 2, \quad
  p_{3} = - (\epsilon_{1} - \epsilon_{2}), \quad
  p_{5} = - (\epsilon_{0} - \epsilon_{2}),
\end{eqnarray}
where $\epsilon_{0}$ and $\epsilon_{1}$ appear from the representation
of ${\cal L}_{(1,1)}$ through the substitution of
Eq.~(\ref{eq:Bruni-17-append}) into Eq.~(\ref{eq:A.25}), and
$\epsilon_{2}$ is an arbitrary parameter that is not determined by
the Leibniz rule for ${\cal L}_{(2,1)}$. 
Thus, the representation (\ref{eq:Bruni-20-append}) is obtained.

%************************************************\\

Finally, we show that the undetermined parameter $\epsilon_{i}$
($i=0,\cdots,3$) can be eliminated through the replacement of
the vector fields $\xi_{(p,q)}^{a}$.
First, we consider the derivative operator (\ref{eq:Bruni-17-append})
of $O(\lambda\epsilon)$.
Then, using Eq.~(\ref{eq:epsilon-1-0-constraint}), the replacement 
\begin{equation}
  \label{eq:xi(1,1)'-xi(1,1)-replacement}
  \xi_{(1,1)}'^{a} = \xi_{(1,1)}^{a} 
  + \frac{1}{2} (\epsilon_{0}-\epsilon_{1}) 
  [\xi_{(1,0)},\xi_{(0,1)}]^{a},
\end{equation}
leads 
\begin{equation}
  \left\{ 
    \frac{\partial^{2}}{\partial\lambda\partial\epsilon} 
    \Phi^{*}_{\lambda,\epsilon}f
  \right\}_{\lambda=\epsilon=0} 
  = 
  {\cal L}_{\xi'_{(1,1)}} f
  + \left(
    \frac{1}{2}{\cal L}_{\xi_{(1,0)}}{\cal L}_{\xi_{(0,1)}} 
    + \frac{1}{2}{\cal L}_{\xi_{(0,1)}}{\cal L}_{\xi_{(1,0)}}
  \right)f .
\end{equation}
This implies that the parameters $\epsilon_{0}$ and
$\epsilon_{1}$ should be chosen so that 
\begin{equation}
 \epsilon_{0}=\epsilon_{1}=\frac{1}{2} 
\end{equation}
without loss of generality. 
Similarly, the replacements of the generators $\xi_{(2,1)}$ and 
$\xi_{(1,2)}$,
\begin{eqnarray}
  \label{eq:replacement-order-2.1}
  \xi_{(2,1)}'^{a} &:=& \xi_{(2,1)}^{a} +
  (\epsilon_{1}-\epsilon_{2})[\xi_{(1,0)},[\xi_{(1,0)},\xi_{(0,1)}]]^{a},
  \\
  \quad
  \xi_{(1,2)}'^{a} &:=& \xi_{(1,2)}^{a} +
  (\epsilon_{1}-\epsilon_{3})[\xi_{(0,1)},[\xi_{(0,1)},\xi_{(1,0)}]]^{a},
\end{eqnarray}
lead to
\begin{eqnarray}
  \left\{ 
    \frac{\partial^{3}}{\partial\lambda^{2}\partial\epsilon}
    \Phi^{*}_{\lambda,\epsilon}f
  \right\}_{\lambda=\epsilon=0} 
  &=&
  {\pounds}_{\xi_{(2,1)}'}
  + 2 {\pounds}_{\xi_{(1,0)}} {\pounds}_{\xi_{(1,1)}'} 
  \nonumber\\
  && \quad
  + {\pounds}_{\xi_{(0,1)}} {\pounds}_{\xi_{(2,0)}}
  + {\pounds}_{\xi_{(1,0)}} {\pounds}_{\xi_{(0,1)}} {\pounds}_{\xi_{(1,0)}},
  \\
  \left\{ 
    \frac{\partial^{3}}{\partial\lambda\partial\epsilon^{2}}
    \Phi^{*}_{\lambda,\epsilon}f
  \right\}_{\lambda=\epsilon=0} 
  &=&
  {\pounds}_{\xi_{(1,2)}'}
  + 2 {\pounds}_{\xi_{(0,1)}} {\pounds}_{\xi_{(1,1)}'} 
  \nonumber\\
  && \quad
  + {\pounds}_{\xi_{(1,0)}} {\pounds}_{\xi_{(0,2)}}
  + {\pounds}_{\xi_{(0,1)}} {\pounds}_{\xi_{(1,0)}} {\pounds}_{\xi_{(0,1)}}.
\end{eqnarray}
These show that the parameters $\epsilon_{2}$ and $\epsilon_{3}$
should be chosen so that 
\begin{equation}
  \epsilon_{2}=\epsilon_{3}=\frac{1}{2}
\end{equation}
without loss of generality. 
Thus, the undetermined parameters in the representation derived
in BGS2003 are eliminated through the replacement of generators
$\xi_{(p,q)}^{a}$, and the representation of the Taylor
expansion in BGS2003 is equivalent to
(\ref{eq:two-parameter-Bruni-30-simpler}) in the main text. 
This also implies that the representation of the Taylor
expansion of the pull-back $\Phi^{*}_{\lambda,\epsilon}Q$ is not
unique, but this non-uniqueness causes no serious problems.

%%%%%%%%%%%%%%%%%%%%%%%%%%%%%%%%%%%%%%%%%%%%%%%%%%%%%%%%%%%%%%%%%%%%%%%
%%%%%%%%%%%%%%%%%%%%%%%%%%%%%%%%%%%%%%%%%%%%%%%%%%%%%%%%%%%%%%%%%%%%%%%

\end{document}